%% file: AGN_FF_maccagni.tex
\documentclass[a4paper,11pt]{article}
\usepackage{aaskaiid}
\usepackage[utf8]{inputenc}
\DeclareUnicodeCharacter{00A0}{ }
\usepackage{tabularx}
\usepackage{makecell}
\usepackage[]{graphicx} % remove 'demo' option in real document
\usepackage{subcaption}
\usepackage{orcidlink}

\newcommand{\HI}{\mbox{H\,{\sc i}}}

\usepackage{printlen}
\usepackage{layouts}

\def\Htwo{H$_{\,2}$}

\newcommand{\kms}{$\,$km~s$^{-1}$}

\newcommand{\Jyb}{Jy\,beam$^{-1}$}

\newcommand{\mJyb}{mJy\,beam$^{-1}$}

\newcommand{\whz}{W~Hz$^{-1}$}
\newcommand{\msun}{{${\rm M}_\odot$}}

\newcommand{\cmsq}{cm$^{-2}$}

\newcommand{\eg}{\mbox{e.g.}}
\newcommand{\ie}{\mbox{i.e.}}

\newcommand{\nhi}{{$N_{\rm H {\hskip 0.02cm \tt I}}$}}

\title{AGN Feeding \& Feedback Over the Galactic Scales }
\ShortTitle{AGN feeding \& feedback}

\author[1,2]{Filippo M. Maccagni\orcidlink{0000-0002-9930-1844}}
\author[3]{Vincenzo Mainieri\orcidlink{0000-0002-1047-9583}}
\author[4]{Isabella Prandoni\orcidlink{0000-0001-9680-7092}}
\author[5]{Massimo Gaspari\orcidlink{0000-0003-2754-9258}}
\author[6,7,8]{W.~J.~G. de Blok\orcidlink{0000-0001-8957-4518}}
\author[9]{Ilaria Ruffa\orcidlink{0000-0002-4578-1205}}
\author[10]{Stanislav S. Shabala\orcidlink{0000-0001-5064-0493}}
\author[11]{Dipanjan Mukherjee\orcidlink{0000-0003-0632-1000}}
\author[12]{Mayur B. Shende\orcidlink{0000-0002-0373-4857}}
\author[13]{Antonino Marasco\orcidlink{0000-0002-5655-6054}}
\author[1]{Paolo Serra\orcidlink{0000-0001-5965-252X}}

\affiliation[1]{ INAF -- Osservatorio Astronomico di Cagliari, via della Scienza 5, 09047, Selargius (CA), Italy}
\affiliation[2]{Wits Centre for Astrophysics, School of Physics, University of the Witwatersrand, 1 Jan Smuts Avenue, 2000, Johannesburg, South Africa}
\emailAdd{filippo.maccagni@inaf.it}
\affiliation[3]{ESO, Karl-Schwarschild-Strasse 2, 85748 Garching bei München,
Germany}
\emailAdd{vmainier@eso.org}
\affiliation[4]{INAF -- Istituto di Radioastronomia, Via P. Gobetti 101, 40125 Bologna, Italy}
\affiliation[5]{Department of Physics, Informatics and Mathematics, University of Modena and Reggio Emilia, 41125 Modena, Italy}
\affiliation[6]{Netherlands Institute for Radio Astronomy (ASTRON), Oude Hoogeveensedijk 4, 7991 PD Dwingeloo, the Netherlands}
\affiliation[7]{Dept.\ of Astronomy, Univ.\ of Cape Town, Private Bag X3, Rondebosch 7701, South Africa}
\affiliation[8]{Kapteyn Astronomical Institute, University of Groningen, PO Box 800, 9700 AV Groningen, The Netherlands}
\affiliation[9]{INAF, Arcetri Astrophysical Observatory, Largo Enrico Fermi 5, I-50125 Florence, Italy}
\affiliation[10]{School of Natural Sciences, University of Tasmania, Private Bag 37, Hobart, TAS, 7001, Australia}
\affiliation[11]{Inter-University Centre for Astronomy and Astrophysics, Post Bag-4, Pune University, Pune - 411007, India}
\affiliation[13]{INAF -- Padova Astronomical Observatory, Vicolo dell'Osservatorio 5, I-35122 Padova, Italy}

\abstract{Active Galactic Nuclei (AGN) are key drivers of galaxy evolution, triggered by cold gas accreting onto a super-massive black hole. However, the processes regulating this gas accretion (feeding) and how AGN alter the interstellar medium to affect star formation (feedback) remain poorly understood. A major observational challenge is the vast range of spatial scales involved: AGN fuelling and jet-ejection occur over the sub-pc scales, while AGN feedback shocks and heats the ISM preventing star formation over the galactic and circum-galactic scales. Moreover, it is unclear how short stochastic AGN episodes are connected with the long timescales of gas accretion and star formation.

In this manuscript, we illustrate how SKAO will provide the unprecedented opportunity to solve the observational limitations of AGN feeding and feedback studies by observing hundreds of nearby AGN down to low radio powers ($10^{21}$~\whz). Simultaneous SKA-Low and Mid observations of nearby galaxies will trace the thermal emission associated with star formation and AGN feedback and the synchrotron emission of their jets of relativistic plasma. These broad-band radio observations enable the detailed characterisation of the AGN duty-cycle, unravelling the time-scales of the nuclear activities.

Reaching in 10 hours neutral atomic hydrogen (\HI) column density sensitivities $\sim 10^{19}$~\cmsq\ at arcsecond resolution, SKA AA4 observations will trace the typical low column density of HI gas in AGN inflows and outflows, to understand the impact AGN feedback over the full galaxy and trace fuelling processes from the environment onto the SMBH. Combining SKA with mm, sub-mm and optical Integral Field Spectrographic observations at comparable arcsecond resolution will provide an exhaustive understanding of the link between multi-phase AGN feeding and feedback processes and star formation.}

\begin{document}
\include{journal-names}

\maketitle

\section{Introduction: The co-evolution of galaxies and their nuclear activity}
\begin{figure}[h]
    \centering
	\includegraphics[width=0.8\columnwidth]{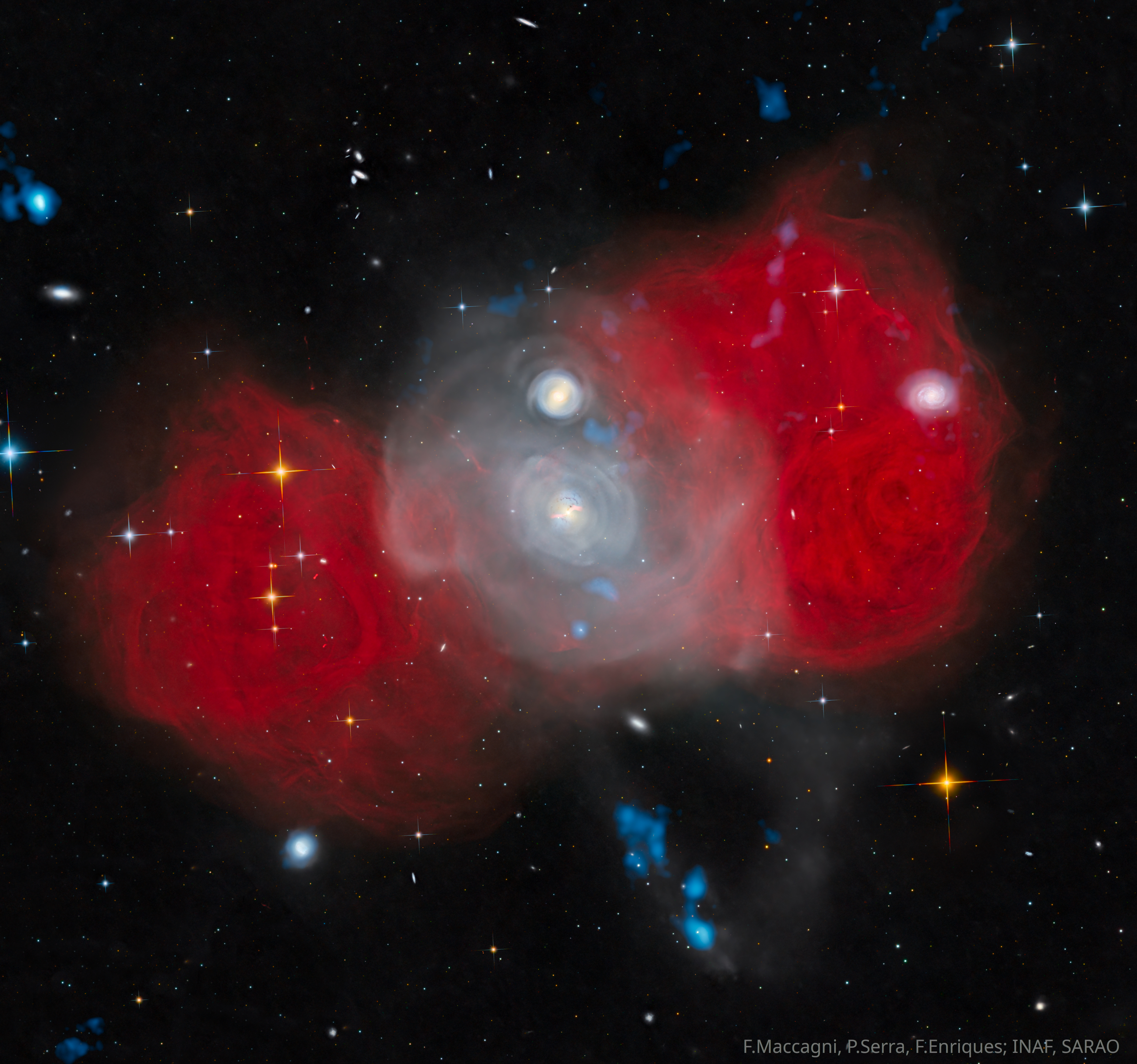}
    \caption{1.4 GHz continuum emission of Fornax A (red) overlaid with the neutral hydrogen clouds (blue) detected in the galaxy and its environment [12,13]. The optical image is a composite of the 63-h exposure taken with LRGB filters with a 14-cm CFF refractor located in Hakos Astrofarm, Namibia, by astrophotographer F. Enriques.}
    \label{fig:fornaxA}
\end{figure}

Observations indicate that accretion of gas onto the super-massive black holes (SMBHs) of galaxies is tightly related with the speed (velocity dispersion) with which the stars in galaxies move, with more massive SMBHs inhabiting galaxies with a higher stellar velocity dispersion~\citep[\eg][]{Kormendy2013}. This unexpected result indicates that the episodes of accretion which occur rapidly and stochastically on scales of less than a hundred pc around the SMBH are connected with the dynamical evolution of galaxies occurring on scales of several kpc. Accretion onto the SMBH generates one of the most energetic phenomena in the Universe, known as Active Galactic Nucleus (AGN), for which we show one of the most nearby examples in Fig.~\ref{fig:fornaxA}. AGN affect the star formation (SF) of the host galaxy, abruptly and rapidly changing the physical conditions of its fuel reservoir. They can rapidly extinguish SF on a global scale by delaying the infall and cooling of the accreting gas~\citep[\ie\ negative AGN feedback;][]{Harrison2014}, but they can also significantly increase it on local scales by compressing the gas and making it easier to cool, coalesce and form stars~\citep[\ie\ positive feedback;][]{Silk2013}. AGN release their energy either through radiation winds which shock-heat the surrounding gas, or through relativistic outflows of non-thermal plasma (\ie\ radio jets) which shock and mechanically displace the ISM. Even though there is a broad-brush dichothomy between radiative and jetted AGN~\citep[][]{Best2012}, there is a number of AGN producing both powerful radiative winds and jets, see for example, Centaurus A~\citep{McKinley2018,McKinley2022}, Mrk 231~\citep{Rupke2011,Morganti2016}, 3C84~\citep{Pedlar1990,Fabian2011}. AGN feedback typically gives rise to fast multi-phase gaseous outflows ejected from the very proximities of the SMBHs and expanding through the host galaxy out to the circum-galactic medium (CGM)~\citep[][]{tombesi2014}. In this context, some of the key open questions are:
\begin{itemize}
    \item define how outflows are driven from the circum-nuclear to the circum-galactic scale;
    \item determine how feedback displaces and heats the ISM and CGM;
    \item understand how efficiently it impacts ISM-CGM and therefore the formation of new stars.
\end{itemize}

\subsection{The AGN feeding and feedback loop} 
The presence of winds and jets in AGN is linked to the fuelling of the SMBH. The radiative output of AGN increases with the efficiency of gas accretion onto the SMBH while the mechanical power of jet diminishes, even though jets are generated across the full spectrum of accretion efficiencies~\citep[][]{Sadowski2017}. Radio sources with low radiative output (low-excitation radio galaxies, LERGs) are the dominant population of radio galaxies in the local Universe and release their energy almost entirely mechanically through their jets. Even though reservoirs of cold gas are often found in their innermost kilo-parsec~\citep[][]{Onishi2017,Ruffa2019,Temi2022}, it is still unclear how this gas has been brought from the IGM, through the galaxy, into the the central regions. The environment of AGN plays a crucial role in their feeding: in the centre of clusters the cooling of the hot intracluster medium funnels cold gas filaments and clouds onto the central regions of the brightest cluster galaxy and often triggers an AGN with extended radio jets~\citep[\eg][]{Tremblay2018,Russell2017,McDonald2019,Tamhane2022,Olivares2022,olivares2025hx-ray-afd,Oosterloo2024}. Less clear is how jets are triggered in galaxies living in poorer environments, where the gravitational potential is shallower~\citep[][]{ruffa2019b}. AGN fuelling can be of external origin where gas is brought into galaxies by mergers or tidal interactions, or internal where turbulence in the hot halo leads the formation of cold gas clouds which rain onto the SMBH~\citep[Chaotic Cold Accretion, CCA][]{Gaspari2013} or bars in the galactic bodies driving torques within the circum-nuclear regions~\citep[][]{Combes2013,Combes2014}. On top of that, gas-rich minor mergers (as traced by galaxies with dust-lanes) had a high AGN fraction, suggesting a link to fuelling via gas-rich minor mergers~\citep[][]{Shabala2012}.

Optical and radio observations~\citep[\eg][]{Woltjer1959,Marconi2004,Best2005} showed that SMBHs are characterized by recursive episodes of accretion, and that the AGN undergo a self-regulated feeding and feedback loop. The gas that fuels the SMBH gets regularly heated and stops being accreted, typically after a few (hundred) Myr, making the AGN flicker for most of the lifetime of the galaxy~\citep[][]{Schawinski2015}.

AGN can go through a feeding and feedback loop, in which the gas that fuels the SMBH is regularly heated by it and, after typically a few hundred Myrs, stops being accreted. After cooling of the gas, accretion may restart, making the nuclear activity flicker for most of the life of the host galaxy~\citep[i.e. AGN duty-cycle;][]{Schawinski2015}.The duty-cycle of AGN seems to be self-sustained by the delicate balance between the  opposite phenomena of feeding and feedback. While jets and radiation pressure shock and heat the surrounding ISM and CGM preventing accretion, they also inject turbulence which may cause cooling from the hot halo. In the same time environmental interactions also cause gas accretion and AGN triggering. Which fuelling mechanisms regulate the AGN duty-cycle is still hotly debated. In this context, some of the key open questions are:
\begin{itemize}
    \item what physical processes set the condensation of hot gas into a multiphase fuel reservoir, and under which conditions does this reservoir form around the SMBH?
    \item how does CCA–driven rain supply mass and redistribute angular momentum to the SMBH over cosmic time, sustaining the observed recursive AGN duty-cycle?
    \item how efficiently do AGN outflows/jets couple to the surrounding medium, and can such self–regulated feedback maintain the halo near quasi thermal equilibrium?    
\end{itemize}

\subsection{The spatial scales of AGN feeding \& feedback} 
\begin{figure}[h]
    \centering
	\includegraphics[width=0.7\columnwidth]{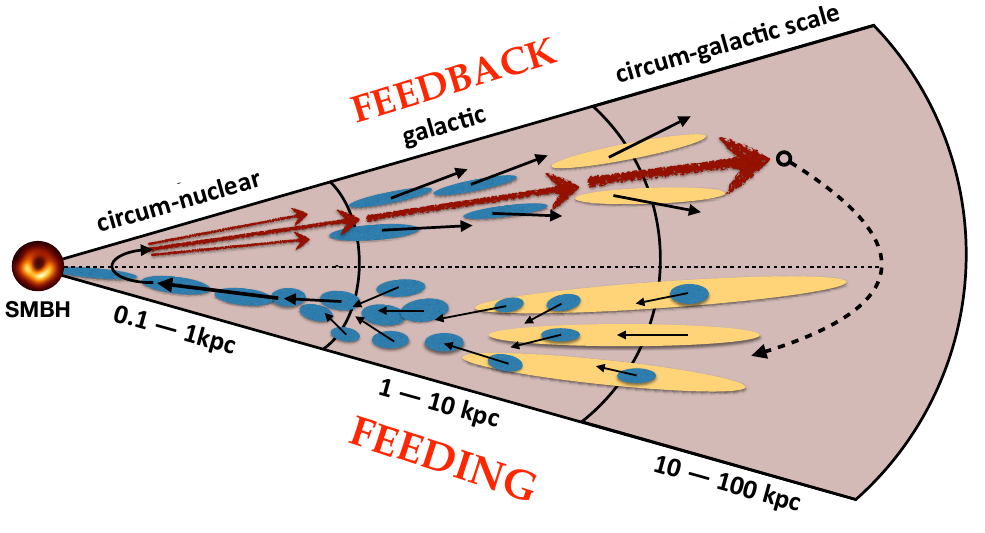}
    \caption{Diagram of key spatial scales tied to AGN feeding and feedback~\citep[reproduced from][review]{Gaspari2020}: the circum-nuclear and galactic (meso-scales) and the circum-galactic (macro-scales).}
    \label{fig:bwhSmall}
\end{figure}

Schematically, AGN feeding and feedback mechanisms occur and co-exist over three main spatial scales~\citep[see Fig.~\ref{fig:bwhSmall}][]{Gaspari2020,Zajacek2022NatAs}. On the circum-nuclear scale (0.1 -- 1 kpc) accretion onto the SMBH occurs and relativistic jets and radiative winds are injected into the ISM. On the galactic scale (1 -- 10 kpc) the jets and winds carve their way through the galactic body~\citep[\eg][]{Cicone2014,Oosterloo2017,Venturi2021}. In the outskirts and environment of the galaxy (circum-galactic scale, 10 -- 100 kpc) AGN jets inflate gigantic lobes of relativistic particles~\citep[\eg][]{Fabian2012}, and galaxy interactions, ram pressure and condensation of the hot halo may trigger the gas to fall back into the galaxy and (re-)start the AGN~\citep[\eg][]{Storchi2019}. In the CCA/black hole `weather' framework (Fig.~\ref{fig:bwhSmall}), these three regimes are tightly coupled: turbulence and cooling in the extended hot halo (macro) seed multiphase condensation on galactic scales (meso), where cold filaments and clouds form and partially rain inward, while on sub-pc scales (micro) this chaotic rain feeds the SMBH and powers jets/winds that reshape the larger-scale atmosphere.

Connecting the small scales of the circum-nuclear regions with the circum-galactic environment is observationally challenging. Across all wavelengths, astronomical instruments with high resolution ($\sim 1''$) have typically small fields of view ($\lesssim 10'$), and vice-versa. Recently, mm and sub-mm arrays like ALMA and Integral Field Spectrographs (IFS) such as MUSE have carried out high resolution studies of the molecular and ionised ISM of nearby AGN, revealing that gaseous outflows extend beyond the circum-nuclear regions~\citep[\eg][]{Cicone2014,Oosterloo2017,Murthy2022,Audibert2025,Ruffa2025,Cresci2015,Mingozzi2019,Venturi2021,Venturi2023}, nevertheless their impact over the galactic scales remains debated.

The high dynamic range in spatial scales represents a challenge also in the theoretical context. Most jet simulations, for example, focus on either the micro-scales close to the SMBH~\citep[][]{Bicknell1984,Wagner2012,Mukherjee2018,Young2025}, or on the circum-galactic scales ~\citep[][]{English2016,Yates2023,Jerrim2025}.

\subsection{Multi-phase AGN feedback}

Several observations have shown that all gas phases (cold neutral atomic, molecular and hot-ionised) of the ISM are involved in AGN feedback phenomena and span a very broad range of physical scales~\citep[\eg][]{Rupke2005,Alatalo2011,Alatalo2015,Oosterloo2017,Oosterloo2024}.
Observationally, outflows are commonly detected in AGN. X-ray and UV emission and absorption line studies revealed outflows with velocities of thousands km/s or even higher, both at low and high redshift, both on pc scales~\citep[][]{Weymann1991,Crenshaw1999,Chartas2002} and on kpc-scale via high-resolution IR and mm spectroscopic studies~\citep[\eg][]{Alexander2010,Feruglio2010,Sturm2011,Rupke2011,Cicone2014,ForsterSchreiber2014,Harrison2014,Genzel2014,Perna2015}. In detail, the outflows ejected at ultra-fast velocities ($\gtrsim 10^3$~\kms) in the proximity of the SMBH \citep[so called UFOs, \eg][]{Tombesi2013} expand through the galaxy shocking the surrounding ISM. Their energetics remain strictly connected with the energetics of the outflows onto the galactic scales~\citep[][]{Fiore2017,Smith2019} under two distinct regimes: momentum-driven and energy-driven. Momentum-driven outflows rely on the pressure of the winds and jets expanding through the medium, while in energy-driven outflows the radiation of the AGN shocks and heats the ISM which expands adiabatically through the galaxy. The momentum rate of energy-driven outflows is expected to be higher than the one of momentum driven outflows~\citep[][]{King2015,Costa2020}. Hence, energy driven outflows may entrain higher outflow rates and generate efficient feedback which affects the SF of the galaxy as a whole. Nevertheless, it is unclear how AGN outflows are driven throughout the galaxy, because obtaining simultaneous measurements of the outflows on both the circum-nuclear and circum-galactic scales is challenging.

The measurement of the momentum of the galactic outflow may be biased by not being able to recover the full mass of the  outflows. This is supported by theoretical models~\citep[][]{Faucher-Giguere2015,Costa2020}, for which all outflows are likely energy driven, but part of the outflowing mass could be missed. Observations show that the bulk of the outflows appears to be in the cold phase of the ISM~\citep{Cicone2014,Fiore2017,Fluetsch2019,Murthy2025}. In some cases, when the molecular or atomic masses of the outflows have been measured over the galactic scales~\citep[IRAS 1790+454, Mrk231, I Zw 1, IRAS F50189-2524,APM 08279+5255][]{Morganti2016,Feruglio2017} the outflows appear to be energy-driven, and thus could potentially efficiently shock-heat the ISM to interrupt SF. Nevertheless, in most cases cases, such as nearby Seyfert \mbox{IC~5063}, when the outflow masses of all phases are measured~\citep[][]{Morganti1998,Morganti2015}, the outflow seems to be momentum-driven, and the sole expansion of the radio jets is sufficient to carry the outflow throughout the galaxy~\citep[see for example, Fig. 9 and Fig. 17 of ][respectively]{Marasco2020,bonanomi2023another-15f}. Momentum-driven outflows could also have a higher impact than expected, given that outflowing gas is sometimes detected in the direction perpendicular to the expanding jets~\citep[\eg][]{Venturi2021,Audibert2025}, thus increasing the efficiency of feedback. Some detailed hydrodynamical models of jet expanding through the medium support this scenario and indicate that, depending on the jet power and their orientation with respect to the clumpy medium, radio jets may produce efficient feedback~\citep[\eg][]{Wagner2012,Mukherjee2018,Meenakshi2022,Mukherjee2025}. 
A definite answer on the mechanisms driving AGN outflows and the efficiency of their feedback can be only given by tracing the cold phases of the outflows (molecular and atomic) from the circum-nuclear to the circum-galactic scales at high resolution. This will enable us to localise the bulk outflowing gas, to accurately measure its mass, velocity and energetics and to determine how efficiently the energy ejected by the SMBH changes the physical conditions of the ISM throughout the whole galaxy.

In this context, the major questions for galaxy-scale feedback are: 
\begin{itemize}
\item How is the energy ejected by the AGN distributed between heating the ISM, blowing it out of the galaxy and injecting turbulence in the CGM? - and how does this change for jet and environment properties. It is important to know where and how the feedback is imparted, because energy can be “wasted” and the gas would cool and return rapidly onto the galaxy.
\item How are AGN most effective in disrupting the surrounding ISM over the galactic scales, through radiation pressure, low or high power radio jets? Recent theoretical work suggests that jet power makes a big difference: low-power jets do not sufficiently shock the gas to drive ionised outflows because it recombines and cools again quickly, whereas high power jets should drive ionised outflows~\citep[][]{Perucho2024}. 
\item What is the connection between outflows and maintenance-mode feedback~\citep[\ie\ their efficiency `keeping the haloes hot', \eg][]{Ciotti2010} and what is the role of radio jets?~\citet{Young2025} show that galaxies in isolated environments are more likely to look asymmetric (one lobe longer than the other) on scales of $\sim$~20 kpc – because of the different paths the two jets take through the clumpy ISM within the host galaxy. So the shorter lobe will have had more feedback imparted by the jet on smaller ($\lesssim 1$ kpc) scales.
\end{itemize}

% {\bf Main open question of AGN feedback: estimate the efficiency of AGN outflows in changing the physical condition of the ISM and CGM and change significantly the star formation history of their host galaxy}

\section{Neutral Hydrogen \HI\ in AGN}
\label{sec:histudies}

Observations of the neutral atomic hydrogen (\HI) are excellent at tracing the complex cycle of gas in AGN feeding and feedback phenomena, over all spatial scales. \HI~low-density clouds detected in absorption (\nhi$\leq 10^{20}$~\cmsq) have been associated with both inflows onto the SMBHs and AGN outflows driven by jets, but also winds~\citep[see][ for a census]{Morganti2018} over the circum-nuclear and galactic scales. However, absorption studies remain strongly hampered by the distribution of the underlying radio continuum, and are therefore very limited in determining the extent of the detected inflows and outflows, while the low-column density HI of outflows and gas accretion ($\lesssim 10^{19}$~\cmsq) has remained so far unaccessible to emission studies. \HI\ also traces the bulk of the cold gas in disks, and has also been a prime tracer of environment-driven galaxy evolution allowing us to study several environmental phenomena, such as tidal interactions, mergers, ram pressure, condensation induced by turbulence~\citep[][]{Chung2007,Wang2020,Maccagni2014}. Neutral atomic hydrogen studies are hence ideal to trace the impact of AGN from the circum-nuclear to the circum-galactic scales and to identify which mechanisms trigger and fuel nuclear activity.

\subsection{Past \HI\ studies in AGN}
Before the SKA pathfinders and precursors, the sensitivity and resolution of past radio-interferometers limited the studies of the low-column density \HI\ involved in AGN feeding and feedback processes mainly to absorption studies of the innermost circum-nuclear regions~\citep[see][]{Mahony01.2026.SKA} or to low-resolution ($\gtrsim 1'$) \HI\ emission studies of the circum-galactic scales. Nevertheless, several phenomena of AGN feeding and feedback have been studied with \HI\ observations.

For example, in Centaurus A, the closest AGN to us ($D=3.4$ Mpc), the outer shells of \HI~\citep[][]{Schiminovich1994} are perturbed by the expansion of the radio jet~\citep[][]{Oosterloo2000}, and in their proximity the pressure of the jet ignited a new region of SF~\citep[][]{Crockett2012,Santoro2016,Santoro2018,salome2024}. In the Minkowski Object, SF has been ignited by the expansion of the jets of companion galaxy NGC541 through the IGM. In that object, the \HI~traces the gas cooling in the IGM because of the shocks generated by the jets~\citep[][]{Croft2006}.

\HI\ absorption showed the first evidences of cold gas inflows in early type galaxies~\citep[][]{vanGorkom1989}, as well as in the brightest galaxy of clusters~\citep[][]{taylor1999} and detected in Seyfert galaxies several circum-nuclear disks tracing the cold-gas reservoirs of their nuclear activity~\citep[\eg][]{Gallimore1999}. 

\begin{figure}[ht]
    \includegraphics[clip,width=0.48\textwidth]{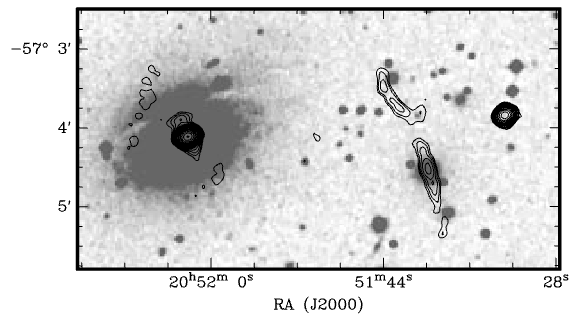}
    \includegraphics[clip,width=0.48\textwidth]{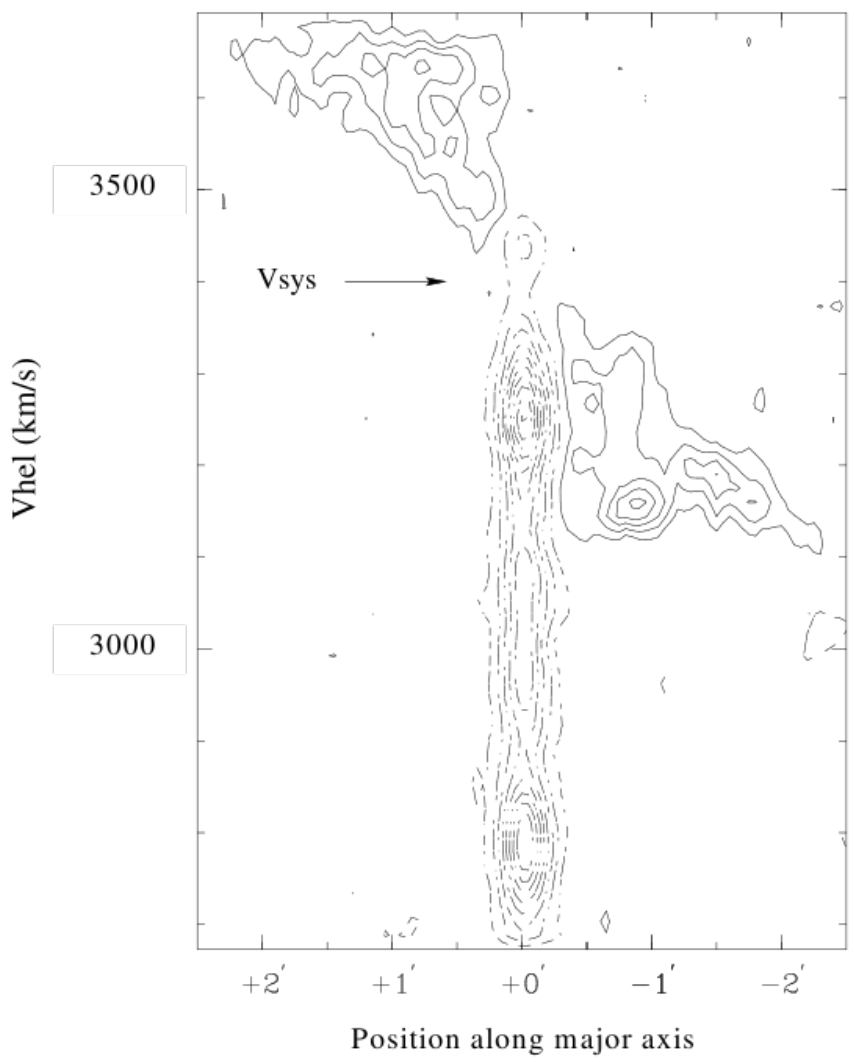}
    \caption{{\em Left Panel}: ATCA 1.4 GHz radio continuum image of \mbox{IC~5063} (source on the left) superposed on a DSS image. Contour levels range from 2.5~\mJyb to 1.16~\Jyb, in steps of a factor 1.5. {\em Right Panel}: Position velocity diagram taken along the major axis (PA $120^\circ$) of the disk of \mbox{IC~5063}~\citep[Figures 2 and 5 of][]{Morganti1998}.}
    \label{fig:ic5063}
\end{figure}

Broad blue-shifted wings in the \HI\ absorption lines can trace gaseous outflows entrained by the AGN radio jets (most commonly), but also radiative winds. Some  examples of \HI\ outflows in AGN are 3C12.50~\citet{Morganti2013}, 3C293~\citet{Mahony2013,Mahony2016}, NCG1266~\citet{Alatalo2011}, or see~\citet{Morganti2018} for a full review. Given that most gas involved in feeding and feedback mechanisms is in the cold phase~\citep[e.g.][]{Fluetsch2019,Murthy2022}, potentially, \HI~absorption provides one of the strongest evidences of AGN jets accelerating and clearing gas off a galaxy, changing its star formation history. A prototypical and historical example of a blue-shifted line tracing a cold gas outflow is the \HI~absorption detected against the radio jets of \mbox{IC~5063}~\citep[Fig.~\ref{fig:ic5063}, left][]{Morganti1998}. The right panel of Fig.~\ref{fig:ic5063} shows that the absorption line (dashed lines) extends for $\sim 500$ km s$^{-1}$ beyond the rotation of the galactic disk (traced by \HI\ emission, solid lines). Follow-up VLBI observations~\citep[][]{Oosterloo2000} showed that the outflow is entrained by the expansion of the radio jets, while the Australia Compact Array Observations (ATCA) did not have the spatial resolution to understand if the jet affects the gas also over the galactic scales, where \HI\ is seen in emission.

\begin{figure}[ht]
    \includegraphics[clip,width=\textwidth]{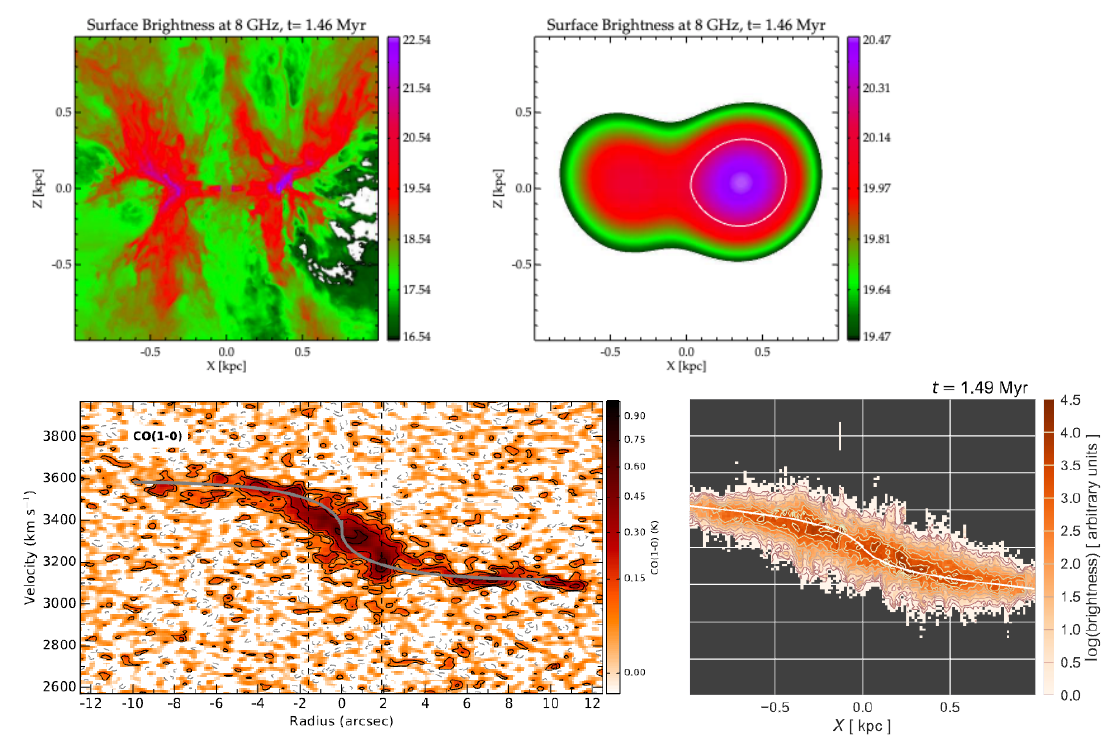}
    \caption{{\em Top Left Panel}: Unconvolved 8 GHz simulated surface brightness profile of a jet with power similar to \mbox{IC~5063} expanding through a galactic disk after 1.49 Myr. {\em Top Right Panel}: Image on the left convolved with a beam of size $234\times234$~pc, as the VLA observations. {\em Bottom Left Panel}: PV diagram along the jet axis of \mbox{IC~5063} of the CO ($1-0$) observed by ALMA. Contour levels are -71.25 (dashed), 71.25 (1.5$\sigma$), 142.5, 285.0 mK~\citep[Fig.~4 of][]{Oosterloo2017}. {\em Bottom right Panel}: PV-diagram along the jet axis of the simulated molecular gas of the current state of~\mbox{IC~5063}~\citep[Figures 7 and 8 of][]{Mukherjee2018}. The direct comparison between the simulations and the cold gas kinematics enables to determine the timescale of expansion of the jets and the mechanical power released into the medium. The dynamic range covered by the colormap (arbitrary units) is set 4.5 to dex, with contours spacing of 0.5 dex.}
    \label{fig:sim_ic5063}
\end{figure}

ALMA and MUSE observations overcame these limitations enabling the high resolution study of the molecular and ionised gas in IC5063. These phases reveal that the outflowing gas not only lies along the direction of expansion of the jets but also perpendicularly to them~\citep[][see also Fig.~\ref{fig:ic5063} (c)]{Morganti2015,Dasyra2016,Oosterloo2017,Venturi2021}. The information on all gas phases involved in AGN outflows enables precise comparison with simulations of jet launching and expansion~\citep[\eg][]{Bicknell1984,Wagner2012,Mukherjee2018,Young2025}, from which is possible to determine the effects of the AGN on the physical conditions of the surrounding ISM. Figure~\ref{fig:sim_ic5063} shows that the kinematics of the cold gas in \mbox{IC~5063} can be directly compared with the ones of the ISM where a low-power jet is expanding through the disk of the galaxy. By measuring the total mass of the outflowing gas and its velocity over the galactic scales, in all phases, and estimating the energetics of the jets it has been possible to assess that AGN feedback in \mbox{IC~5063} is momentum-driven~\citep[see Fig. 9 in][]{Marasco2020}.

\subsection{On-going \HI\ studies in AGN}

\begin{figure}[h]
    \centering
	\includegraphics[width=0.7\columnwidth]{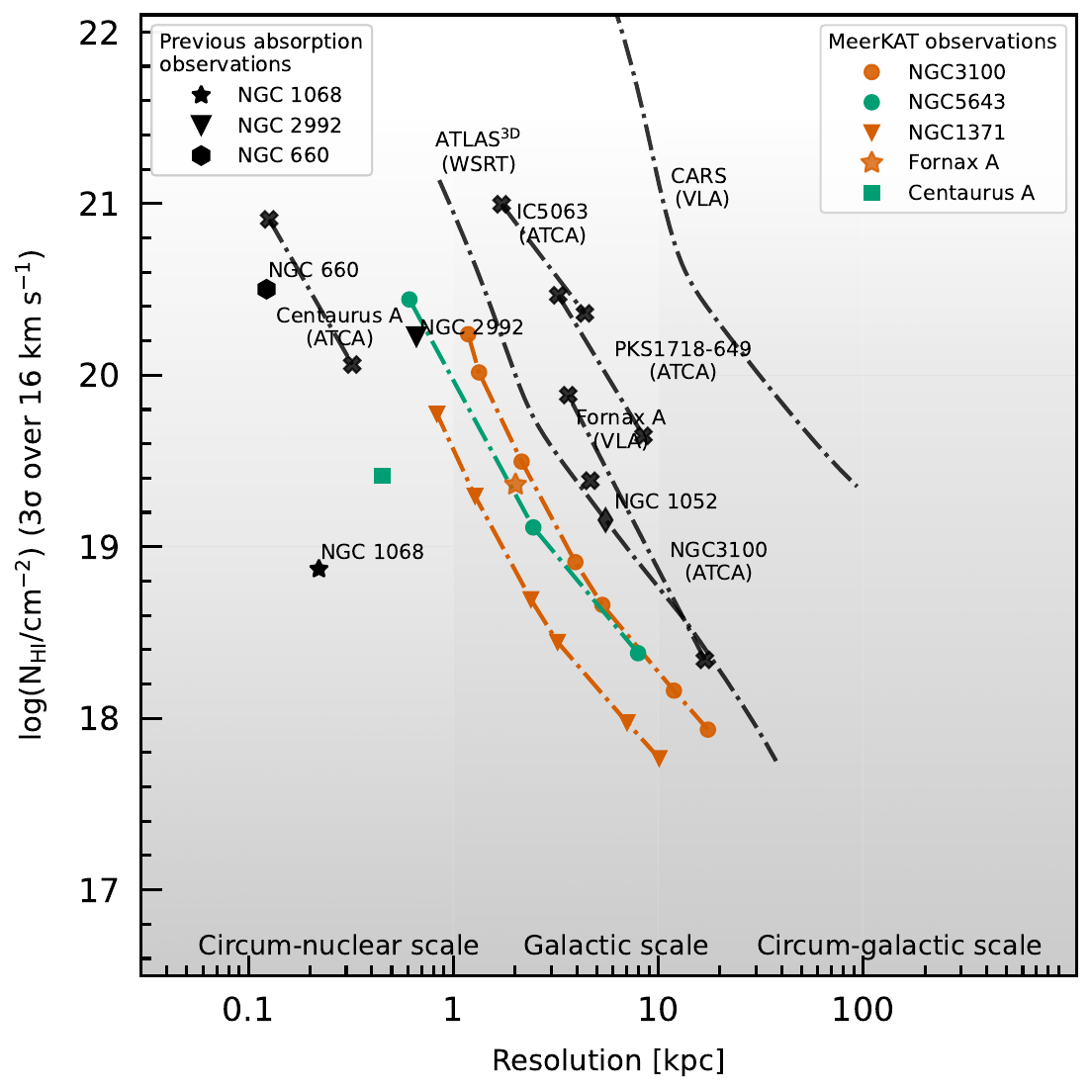}
    \caption{Sensitivity vs. resolution of \HI\ observations of nearby AGN. The black lines and markers show previous \HI\ observations. Orange markers and lines show MeerKAT observations of nearby AGN illustrated in this chapter~\citep[Fornax A, NGC3100 NGC 1371][]{Maccagni2020,Veronese2025}, along with other deep MeerKAT observations shown in green. The MeerKAT observations extend two orders of magnitude into an unexplored regime studying at kpc resolution the low density \HI\ typically involved in AGN feeding and feedback.}
    \label{fig:sensPlot}
\end{figure}

The cold gas involved in AGN feeding and feedback processes can now be observed in great detail by the SKA-Mid precursor MeerKAT in nearby galaxies ($\leq 35$ Mpc). MeerKAT reaches low-column density sensitivities (\nhi$\sim 10^{19}$~\cmsq) with $\geq 20$ hours of observing time and over a field of view of 1 deg$^2$. Figure~\ref{fig:sensPlot}~\citep[adapted from][]{Maccagni2024b} compares the column density sensitivity and physical resolution explored by past \HI\ studies of nearby AGN with recent and on-going \HI\ studies with MeerKAT. The figure shows published \HI\ observations of nearby AGN in emission and absorption (\eg\ IC~5063) with ATCA, the Very Large Baseline (VLA) and the Westerbork Synthesis Radio Telescope (WSRT) and highlights the new parameter space that MeerKAT observations of nearby AGN are exploring. In sources like Fornax A ($D=20.9$ Mpc), NGC~3100 ($D=33.4$ Mpc) and NGC~1371 ($D=22.7$ Mpc). 12-55 hours of MeerKAT observations allow us to investigate the parameter space of low-column density \HI\ over all spatial scales, enabling the study of the typical column densities of \HI\ inflows and outflows (\nhi$\lesssim10^{20}$~\cmsq) with kilo-parsec resolution, or better. So far, the MeerKAT observations shown in Fig.~\ref{fig:sensPlot} are the deepest, high-resolution, high-fidelity \HI\ observations of nearby AGN ever obtained. In the following Sections, we illustrate how these observations, in combination with broad-band continuum observations, shed new light on AGN feeding and feedback studies over the galactic scales. Reaching \HI\ column density sensitivities $\lesssim 5\times 10^{19}$~\cmsq\ with kilo-parsec resolution, or less, we can trace feeding and feedback mechanisms over all spatial scales (from tens of kpc to the sub-kpc scales close the SMBHs) and link the rapid recursive activity of SMBHs to the slow and continuous evolution of their host galaxies.

%MeerKAT, the SKA-Mid precursor, is the one and only instrument capable of detecting low-column density HI over all spatial scales~\citep[][]{Maccagni2024b} and to reveal the complex interplay between AGN and the surrounding ISM and IGM. 

\subsubsection{Fornax A}

The MeerKAT observations of Fornax A (Fig.~\ref{fig:fornaxA}) detected the radio jets and extended lobes ($r\sim 150$ kpc) of this nearby ($D=20.9$ Mpc) AGN with unprecedented detail~\citep[][]{Maccagni2020} and detected, for the very first time, various \HI\ clouds and filaments in its centre and in its environment \citep[in blue in Fig.~\ref{fig:fornaxA};][]{Serra2019,Kleiner2021,Maccagni2021}. Combining the 900 MHz and 1.4 GHz broadband ($\Delta\nu$= 100 MHz) MeerKAT observations with continuum observations between 94 MHz to 217 GHz,~\citet{Maccagni2020} measured the resolved spectral energy distribution of the radio components of the AGN (\ie\ the lobes, the central jets and the core). In the radio spectrum, a relativistic jet continuously injected by relativistic particles is described by a sharp cut-off whose frequency depends on the age of the radiation and its history of injection~\citep[\eg][]{Kardashev1962,Pacholczyk1970,Jaffe1973,Slee2001,Murgia1999,Murgia2011,Harwood2013}. Once the AGN turns off, the spectral shape of the jets is characterized by a new frequency break, after which the spectrum drops exponentially~\citep[\eg][]{Slee2001,Parma2007,Murgia2011}. Measuring the distance between the frequency breaks it is possible to estimate for how long the AGN has been active and for how long it has been in a quiescent phase. In Fornax A, the shape of the spectral energy distribution shows that the AGN is rapidly flickering with the lobes originating by several past ($\gtrsim 12$ Myr) activities, while the central jets did not yet pierce through the galaxy and show a more recent activity ($\sim 6$ Myr) and the core recently re-activated~\citep[also confirmed by VLBI observations][]{Paraschos2024}. 
%The detection of the \HI filaments in the environment of Fornax A revealed that the formation of the radio lobes likely first began $\sim 1$ Gyr ago, when the galaxy underwent a gas rich major merger (Serra et al. 2019, Kleiner et al. 2021). Nevertheless, different feeding mechanisms must be currently on-going to fuel the rapid flickering of the AGN. 

In the centre of Fornax A, neutral atomic, molecular and ionised hydrogen co-exist revealing a complex distribution and kinematics. Even though limited by the coarse spatial and spectral resolution of the \HI\ observations, 20'', 44~\kms,~\citet{Maccagni2021}, performed a combined analysis of the \HI, CO and $H\alpha$ gas kinematics in the innermost 6 kpc of Fornax A. A rapid kinematical diagnostic plot that combines the information from the velocity and dispersion maps of the multi-phase gas in Fornax A~\citep[\ie\ the kinematical plot, k-plot;][]{Gaspari2017,Maccagni2021} allows to identify where the gas is out-flowing, possibly entrained by the radio jets, and where instead is condensing and flowing towards the SMBH. 

The combination of high sensitivity multi-wavelength observations reveales that in Fornax A, feeding and feedback events co-exist in space and time suggesting, they self-regulate the rapid flickering of the nuclear activity. Nevertheless, the analysis of the multi-phase gas is limited by the the coarse spatial and spectral resolution of the L-band observations, being 20 times lower than the MUSE observations of the ionised phase. While the kinematics of the gas indicate that CCA is driving gas from the innermost 6 kpc onto the SMBH, only by studying all phases of the gas at the same comparable resolutions will be possible to precisely assess the physical properties (\ie\ mass, turbulence, cooling time) of the inflowing gas. In particular, the coexistence of extended, turbulent filaments, compact cold clouds, and inflow signatures along multiple position angles is exactly what is expected in a developed CCA rain, where condensation and accretion proceed chaotically rather than through a single coherent disk~\citep{Gaspari2013,Gaspari2017,Wang2023}. On top of that, only by precisely assessing the total out-flowing gas masses throughout the whole galaxy will allow us to determine the feedback energetics and efficiency in unsettling the ISM. Further information on expected SKA-Mid AA4 observations of \HI\ gas in AGN environments is given in Sect.~\ref{sec:cca_synt}.

\subsubsection{\mbox{NGC~3100}}

\begin{figure}[tbh]
    \centering
	\includegraphics[width=0.8\textwidth]{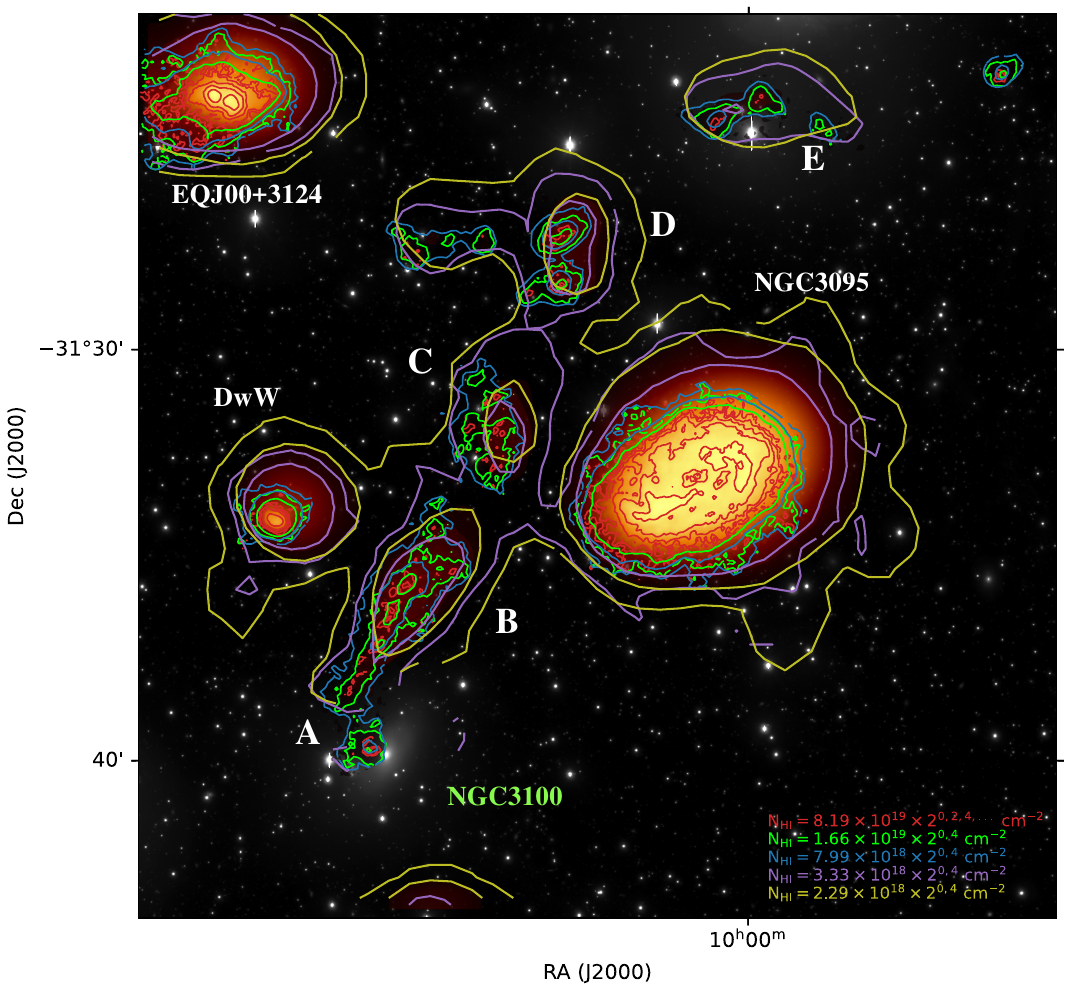}
    \caption{Flux-density \HI\ emission (in orange) detected by MeerKAT, overlaid with a 20' inset of the deep ($27$ mag arcsec$^{-2}$) {\em g,r} optical image from VEGAS~\citep[][]{Maccagni2023}. The \HI\ emission is a composite of flux-density maps derived from the multi-resolution datacubes between $10''$ (red) and $92''$ (yellow-green). A `dark' 300 kpc-long \HI\ filament connects the circum-nuclear regions of \mbox{NGC~3100} with its turbulent circum-galactic environment.}
    \label{fig:ngc3100}
    % \begin{figure}[tbh]
%     \centering
%     \caption{}
%     \label{fig:\mbox{NGC~1371}}
% \end{figure}
\end{figure}

\begin{figure}[ht]
    \includegraphics[clip,width=\textwidth]{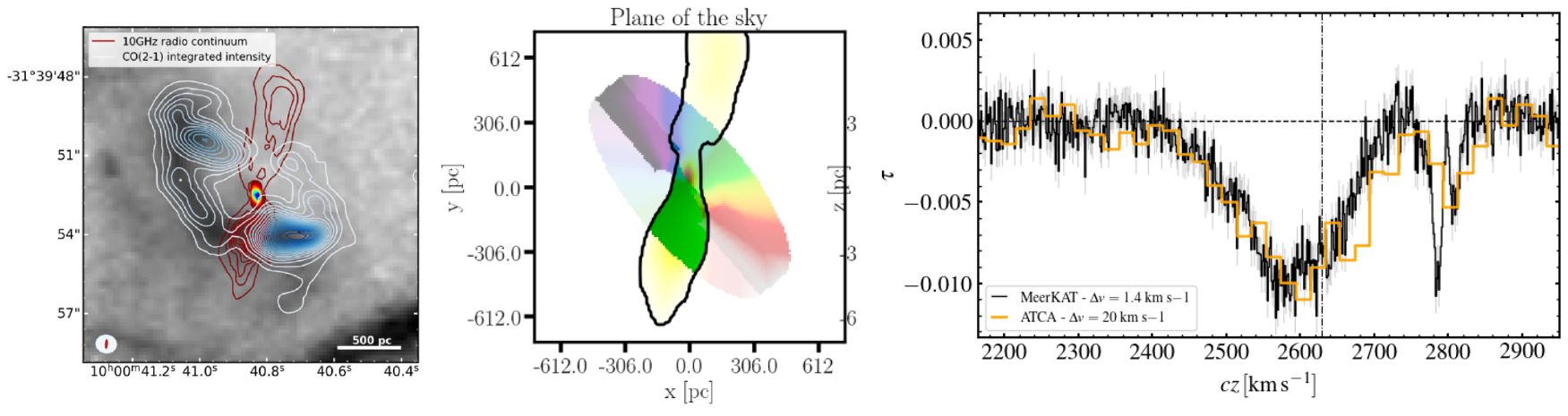}
    \caption{{\em Left Panel}: HST image of the centre of NGC3100 with ALMA CO(2–1) intensity contours, and radio continuum contours from JVLA data at 10 GHz~\citep{Ruffa2019}. {\em Centre Panel}: Kinematical absorption model of the \HI\ disk oriented as the molecular gas disk which best reproduces the HI absorption line detected by ATCA, shown in the right panel. {\em Right Panel}: \HI\ absorption detected in the centre of NGC3100 by ATCA (orange) and MeerKAT (black). The MeerKAT 1.4~\kms\ resolution enables us to spectrally resolve the multiple HI clouds fuelling the AGN~\citep[adapted from ][]{Maccagni2023}.}
    \label{fig:n3100}
\end{figure}

Fornax A belongs to a complete volume and flux limited sample of nearby galaxies classified as LERGS~\citep[$\lesssim 10^{25}$~\whz][]{Ekers1989,Ruffa2019}, which represent the dominant population of jetted-AGN. Even though hosted by gas-poor early-type galaxies, ALMA observations revealed that the circum-nuclear regions of these AGN are rich of molecular gas ($10^7$--$10^9$~\msun) often assembled in regularly rotating disks~\citep[][]{Ruffa2019}, with small kinematical deviations possibly tracing jet-ISM interactions~\citep[][]{Ruffa2022}. ALMA, however, only probes the central few kiloparsecs of these AGN, while much larger-scale imaging is necessary to probe their circum-galactic regions, identify interactions with nearby galaxies, understand which mechanisms drive the gas from the CGM and ISM to the SMBH, under which timescales, and ultimately fully assess the origin of the fuel reservoir of these AGN. Figure~\ref{fig:ngc3100} shows a multi-filter $20'$-wide optical image of another AGN of this complete sample of LERGs, \mbox{NGC~3100} ($D=33.4$ Mpc). The multi-colour contours show the flux density distribution of the \HI\ detected by MeerKAT at resolutions between 10 and 90 arcseconds (1.8 and 16 kpc resepectively). Because of the large field of view and exceptional sensitivity ($5\times10^{18}$~\cmsq\ at $30''$) MeerKAT reveals that a 300 kpc long \HI\ filament extends through two neighbouring galaxies of \mbox{NGC~3100}, connecting the CGM with its innermost regions. Previously, combining ATCA and deep optical observations~\citep[][]{Maccagni2023} it was possible to detect part of the \HI\ filament, determine the absence of an associated stellar counterpart and detect, in absorption, the \HI\ counterpart of the \Htwo\ circum-nuclear disk (Fig.~\ref{fig:n3100}, left). Now, the MeerKAT observations enable us to determine the total mass of the filament and study its kinematics to understand how it formed through the past interactions within the group and under which mechanism it is fuelling the circum-nuclear disk of \mbox{NGC~3100}. Against the radio jets of NGC 3100, \HI\ is detected in a complex absorbing system~\citep[Fig.~\ref{fig:n3100}, right][]{Maccagni2023}. The central panel of Fig.~\ref{fig:n3100} shows that modelling an \HI\ disk with same orientation and kinematics as the molecular gas disk is possible to determine that the bulk of the \HI\ line traces the same circum-nuclear disk, while the redshifted feature does not. Figure~\ref{fig:n3100} also shows the power of high spectral resolution ($1.4$ km s$^{-1}$) observations in fully recovering the information beneath an absorption line. Because of the coarser ATCA spectral resolution the redshifted feature appeared broad and of low optical depth, while new MeerKAT observations (Maccagni et al. in prep.) show that the redshifted feature is a complex of narrow lines tracing single different clouds with high optical depth falling onto the SMBH from within the circum-nuclear disk. These observations have similar spectral resolution to the ones SKA-AA4 will provide. The combination of HI absorption with high resolution molecular gas observations is crucial for the correct interpretation of several \HI\ absorbing systems at low and high redshift~\citep[see for example, PKS 1718-649 and PKS B1740-517]{Maccagni2014,Maccagni2016,Maccagni2018,Allison2015,Allison2019,Mahony01.2026.SKA}. 

The joint study of the atomic and gas phases over the galactic scales of \mbox{NGC~3100} is limited by the $10''$ resolution, which allows to detect the \HI\ in emission over the galactic scales with only one resolution element. Only SKA-Mid AA4 observations will resolve the \HI\ gas in nearby AGN at arcsecond resolution allowing accurate joint studies of all phases of the gas (atomic, molecular and ionised) and direct comparison with hydro-dynamical simulations of the SMBH black hole accretion and jet ejection processes.

\subsubsection{\mbox{NGC~1371}}
\label{sec:ngc1371}

\begin{figure}[tbh]
    \centering
 	\includegraphics[width=0.7\textwidth]{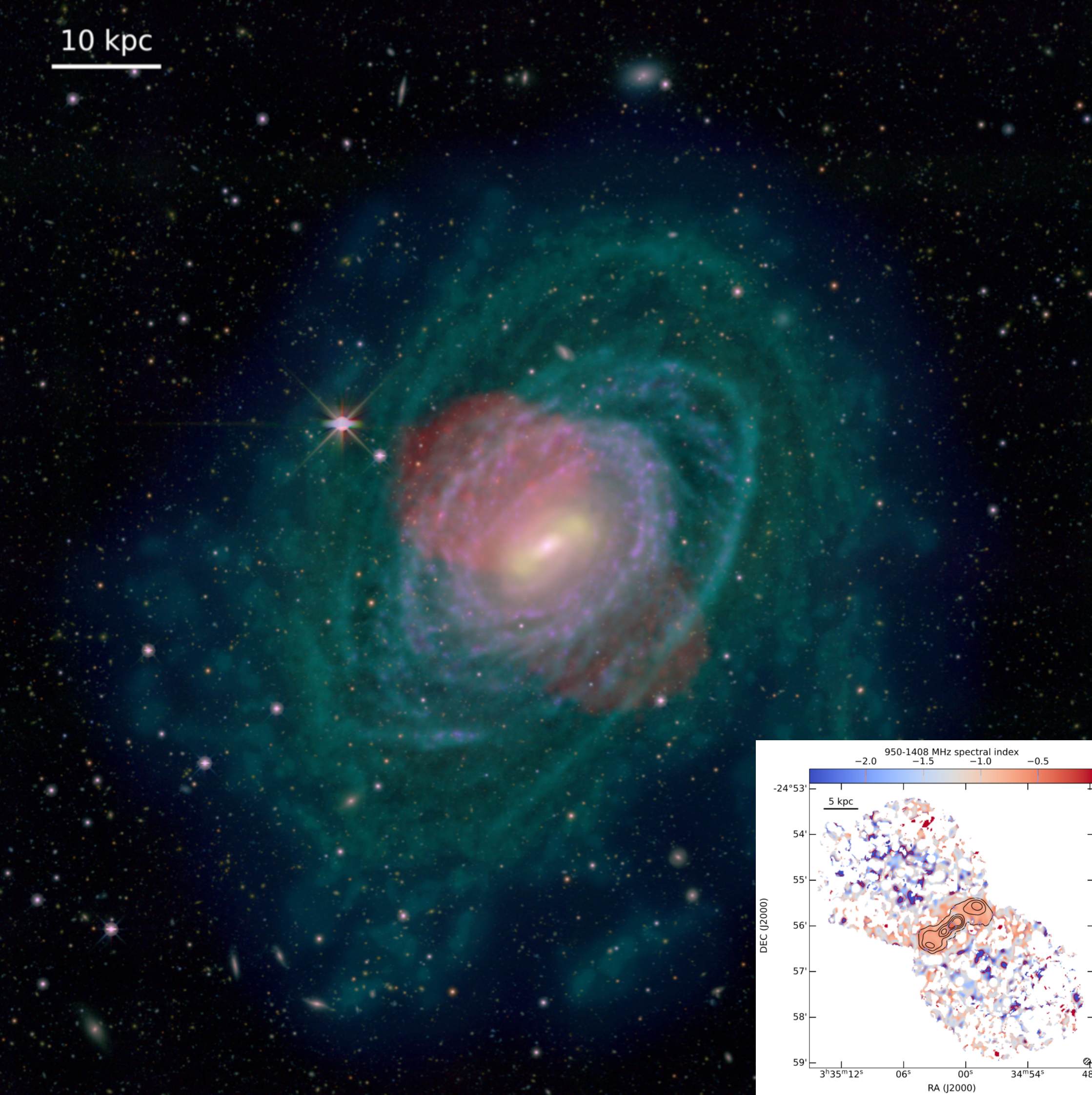}
    \caption{Multiwavelength image of NGC 1371. The background shows the combined {\em gzri} optical image from Dark Energy Camera Legacy Survey~\citep[DECaLS,][]{Dey2019}. The MeerKAT high-resolution 1.4 GHz radio continuum is shown in yellow and red. The UV emission as observed by GALEX is overlaid in pink. The multi-resolution \HI\ from the MHOONGOOSE observations is given in green and blue. The box in the bottom right corner shows the L-band spectral index map of the jets and the lobes~\citep[Fig.~1 of ][]{Veronese2025}.}
    \label{fig:ngc1371}
    % \begin{figure}[tbh]
%     \centering
%     \caption{}
%     \label{fig:\mbox{NGC~1371}}
% \end{figure}
\end{figure}

The MeerKAT 1.4 GHz observations of \mbox{NGC~1371} are another seminal example of how the SKA precursor opens a new space of exploration in the studies of AGN feedback in low-power sources~\citep[$P_{1.4 \rm \, GHz}=3\times10^{21}$~\whz,][]{Veronese2025}. \mbox{NGC~1371} has been observed for 55 hours as part of the MHONGOOSE survey~\citep[][]{Blok2024} which has the science goal to understand how the \HI\ fuels SF in these galaxies and to detect low-column density \HI\ $\sim 5\times10^{18}$~\cmsq accreting onto its disk. The 55hrs L-band continuum observations revealed in detail large-scale radio bubbles (10 kpc) expanding perpendicularly to the disk (Fig.~\ref{fig:ngc1371}). The origin of these bubbles is uncertain,~\citet{Veronese2025} suggest that they may be the back-flow bubbles of the low-power jets ($P_{1.4 \rm GHz}=3\times10^{21}$~\whz) expanding through the disk. The jets may have contributed in clearing the gas out of the innermost regions of this disk, which appear void of both \HI\ and molecular gas, but the role of the radio bubbles, how they formed and if they also contribute to feedback is unknown. Broad-band spectral index studies (rather than in-band, as shown in the box of Fig~\ref{fig:ngc1371}) will be crucial to determine the timescales of the bubbles and link them to the secular evolution of the gaseous disk.  Interestingly, these bubbles have been detected in a handful of AGN in nearby spirals~\citep[see][]{Ledlow1998,Morganti2011,Doi2012,Bagchi2014} but, so far, remained largely unstudied. 

% \begin{figure}[tbh]
%     \centering
% 	\includegraphics[width=0.6\columnwidth]{\mbox{NGC~1371}.pdf}
%     \caption{Multiwavelength image of NGC 1371. The background shows the combined {\em gzri} optical image from Dark Energy Camera Legacy Survey~\citep[DECaLS,][]{Dey et al. 2019}. The MeerKAT high-resolution 1.4 GHz radio continuum is shown in yellow and red. The UV emission as observed by GALEX is overlaid in pink. The multi-resolution \HI from the MHOONGOOSE observations is given in green and blue. The box in the bottom right corner shows the L-band spectral index map of the jets and the lobes~\citep[Fig.~1 of ][]{Veronese2025}.}
%     \label{fig:\mbox{NGC~1371}}
% \end{figure}

\section{Exploring a new parameter space in \HI\ emission studies}

\begin{figure}[h]
    \centering
	\includegraphics[width=0.7\columnwidth]{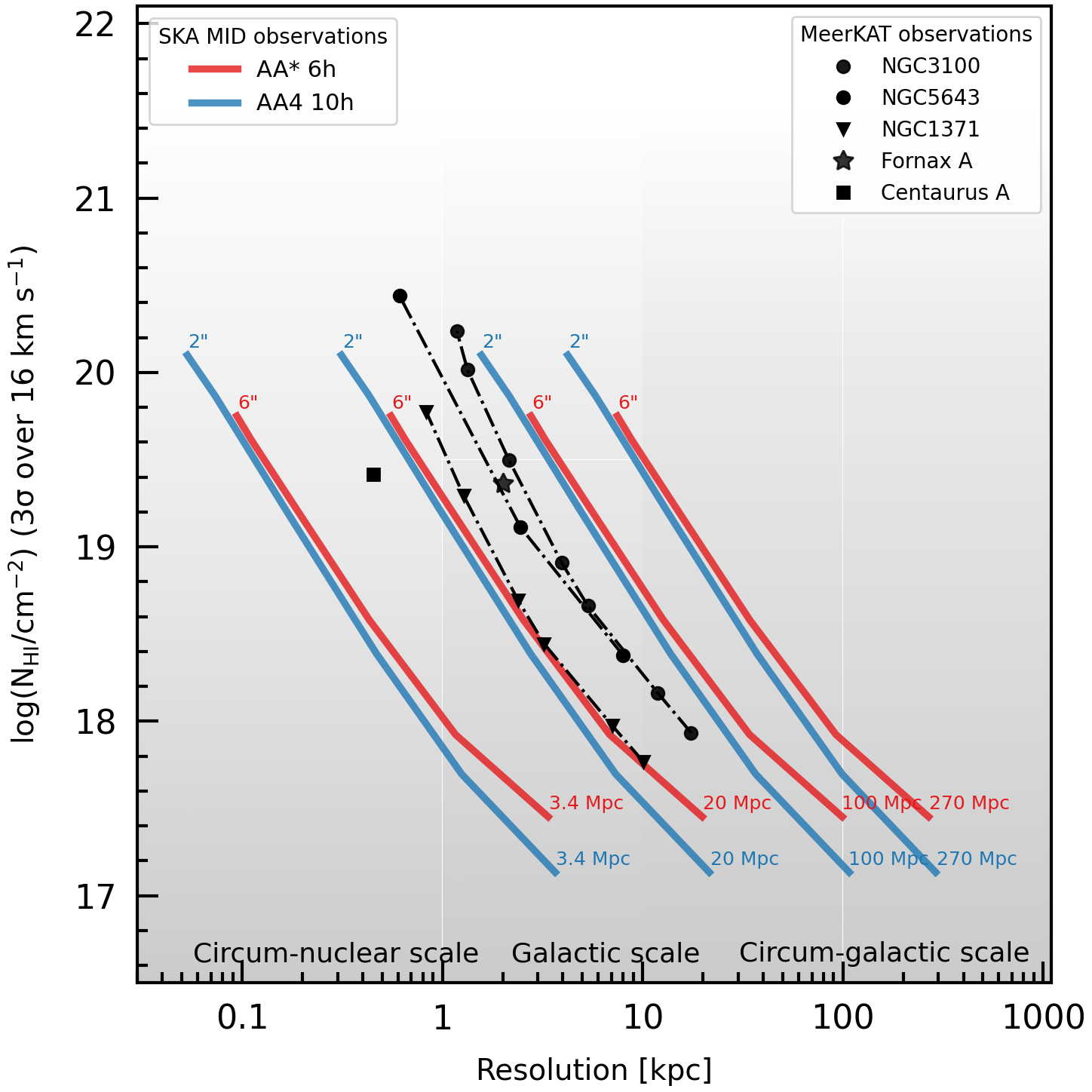}
    \caption{Sensitivity vs. resolution of \HI\ observations of nearby AGN reached by SKA-Mid AA$^*$ and AA4 in six and ten hours of observation, respectively. The AA4 array configuration will allow us to observe the diffuse ($\lesssim 5\times 10^{19}$~\cmsq) \HI\ in Fornax A (20.9 Mpc) at kilo-parsec resolution, and explore the new parameter space of \HI\ $\sim 6 \times 10^{18}$~\cmsq\ with 2-kpc resolution.}
    \label{fig:sensPlot_ska}
\end{figure}

SKA-AA4 will investigate the \HI\ parameter space of sensitivity vs resolution, sensibly improving both. Figure~\ref{fig:sensPlot_ska} shows how with only 10 hours of observations SKA-Mid AA4 will allow us to infer at sub-kiloparsec resolution \HI\ gas at the distance of Centaurus A. SKA-Mid AA4 will observe objects at 20 Mpc distance (as Fornax A) with sensitivity and resolution that MeerKAT can reach only in the very nearby Universe ($\lesssim 10$ Mpc), where there are only a handful of AGN.

The $2''$ spatial resolution will allow us to study the distribution and kinematics of the dense ($\gtrsim 10^{20}$~\cmsq) \HI\ with unprecedented image quality, enabling a direct comparison with mm and IFS observations of the molecular and ionised gas phases. The jet-cold gas interactions will be resolved over the galactic scales in AGN down to low-radio powers ($10^{21}$~\whz), out to 100 Mpc. This will enable a direct comparison with hydrodynamical simulations with physical parameters calibrated to be representative of the observed AGN and their hosts (\eg\ similar SMBH and galaxy mass, ISM gas phases and kinematics, jet power and expansion through the medium). AGN feeding and feedback mechanisms are complex and differ depending on the properties of the AGN and their environment. To fully understand these processes, we must perform detailed jet-HI studies in the whole heterogeneous family of AGN which have jets spanning several orders of magnitudes in power ($P_{\rm 1.4\,GHz}\sim 10^{22-26}$~\whz), different host galaxy type, and environments, from clusters to poor environments. Only by studying such a large range in AGN properties, can we understand AGN feeding and feedback mechanisms, quantify their timescales and measure the effects of jet expansion on the co-evolution of SMBHs and their host galaxy. This will allow us to finally draw solid conclusions on the impact that AGN feeding and feedback loops have on the evolution of their host galaxies.

\subsection{Synthetic observations of AGN feeding with SKA-AA4}
\label{sec:cca_synt}

\begin{figure}[tbh]
    \centering
	\includegraphics[width=0.45\columnwidth]{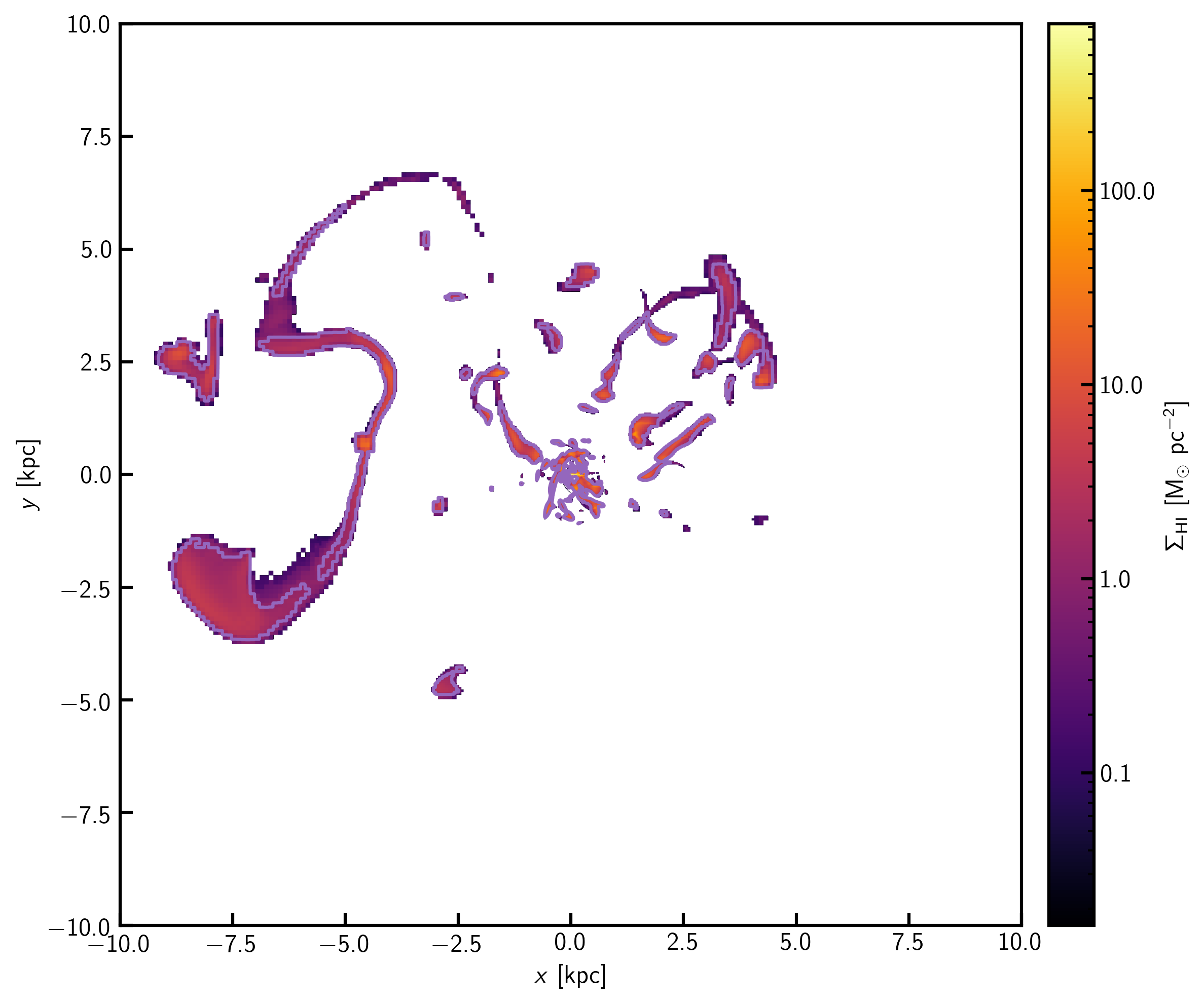}
	\includegraphics[width=0.45\columnwidth]{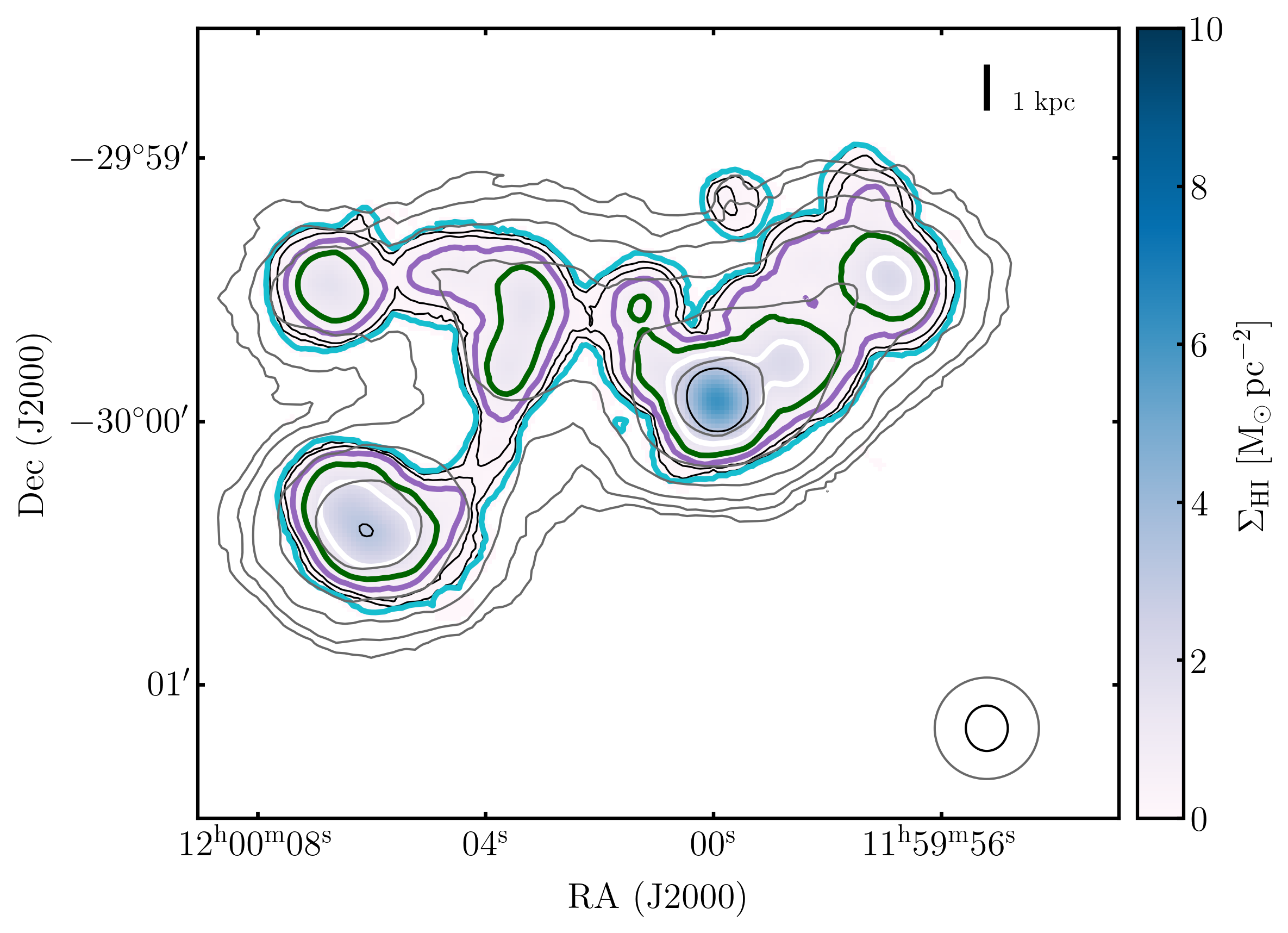}
    \caption{{\em Left Panel}: \HI\ column density map extracted from a simulation of CCA rain (\citealt{Gaspari2017}) in an environment similar to Fornax A (same $M_\star$, SMBH and $M_{\rm ISM}$). The purple contour marks the 0.8~\msun pc$^{-2}$ level, corresponding to an \HI\ column density of $10^{20}$\cmsq.{\em Right panel}: Multi-resolution (10'' and 20'', colored and grey contours, respectively) \HI\ surface brightness map processed from the synthetic 10 hours observations with SKA-Mid AA4 of the \HI\ in the left panel at the distance of Fornax A (20.9 Mpc). Cyan, purple green and white contours correspond to $5\times10^{18}$, $5\times10^{19}$, $1\times10^{20}$~\cmsq, $2\times10^{20}$~\cmsq\ levels, respectively. The 10'' and 20'' PSF ($\approx 1$ and $2$ kpc, rescpectively) are shown in the bottom right corner.}
    \label{fig:cca}
\end{figure}

The analysis of the physical conditions of the \HI\ in Fornax A is limited by the spatial resolution of the L-band observations (Sect.~\ref{sec:histudies}). Fig.~\ref{fig:cca} (a) shows a snapshot of the atomic hydrogen distribution in the innermost 10 kpc of an AGN host with same M$_\star$, M$_{\rm gas}$ and M$_{\rm SMBH}$ as Fornax A extracted from a simulation of CCA \citep{Gaspari2017}, during an active top-down multiphase condensation rain. Since the CCA simulation provides temperature and density of each spaxel but not the \HI\ gas mass per cell, we can derive volumetric \HI\ gas fractions in the simulations from the gas temperature and density using the prescription of ~\citet{Gnedin2014},  which is well suited for elliptical galaxies. A general method to extract \HI\ gas from different simulations (\eg TNG50, FIRE) is illustrated in~\citet{Marasco2025}.

The left panel of Fig.~\ref{fig:cca} shows a synthetic 10 hours SKA-Mid AA4 observation of the \HI\ from the simulation in the left panel. To produce the synthetic observation we generated a visibility dataset of the \HI\ surface brightness of the simulated gas (at the distance of Fornax A, 20.9~Mpc), using a wrapper built on {\tt CASAtool} and {\tt casacore}, where we added the SKA-Low and Mid AA4 array configurations. We processed the mock observation with the same data reduction strategy and tools as the MeerKAT observations~\citep[CARACal, see \eg][]{jozsa2020,serra2023,Blok2024}. This produced a realistic datacube with the noise and clean artefacts expected from a 10 hours SKA-Mid AA4 observation. Running the SoFiA-2 sourcefinder~\citep[][]{westmeier2021} with same parameters as in the MeerKAT observations,  we generated the \HI\ moment maps.

SKA-Mid AA4 observations will enable a direct comparison with the simulations and a detailed study and identification of the gas inflows and outflows. The right panel of Fig.~\ref{fig:cca} shows the \HI\ surface brightness distribution at 20'' and 10'' resolution (approximately 2 and 1 kpc at the distance of Fornax A) as it would be observed by SKA-Mid AA4 in 10 hours. The image quality and spatial resolution at $10^{19}$~\cmsq\ sensitivity (purple contours) is unprecedented and allows to trace the same structures seen at pc resolution in the simulation (left panel). Pushing SKA-Mid AA4 resolution at 2'', we will be able to resolve the high density gas ($\gtrsim 10^{20}$~\cmsq\, green contours) at 20 Mpc with sub-kpc resolution (see Fig.~\ref{fig:sensPlot_ska}). Because of the short observing times (10 hours) SKA-Mid AA4 Band2 surveys will allow to study \HI\ involved in AGN feeding and feedback phenomena over the galactic scales in a representative sample ($\sim$ 1000) of nearby (within $\sim 100$ Mpc) galaxies, thus providing a detailed analysis of the mass kinematics of the cold gas in AGN and a study of the impact of radio jets down to very low-powers ($10^{21}$~\whz). The mock maps based on the CCA simulation already reveal rich feeding–feedback morphology, with a composition of HI filaments and ensemble clumpy clouds, at different meso- and macro-scales.
These SKA-Mid AA4 observations will resolve in detail the physical conditions of the jet–ISM interaction (\eg turbulence, thermal state, and density structure of the gas), enabling CCA diagnostics such as the $k$–plot and condensation ratios (\eg\ $t_{\rm cool}/t_{\rm ff}$, $t_{\rm cool}/t_{\rm eddy}$; \citealt{Gaspari2018}) to directly probe the AGN feeding–feedback cycle. The same diagnostic framework can then be applied statistically in wide SKA surveys over $\sim 1000$ deg$^2$, which will map the radio AGN population at higher redshifts ($z>0.1$).  

% \begin{figure}[h]
%     \centering
% 	% \includegraphics[width=0.45\columnwidth]{P5e43_nw300_0150.png}
%     \caption{{\em Left Panel}: Simulation of 1.4 GHz synchrotron surface brightness produced by the expansion of a low-power jet through a clumpy medium (courtesy of D. Mukherjee). This simulation replicates the radio jets and lobes of \mbox{IC~5063}~\cite[][]{Morganti1998,Mukherjee2018} {\em Centre Panel} Synthetic observation with 10hrs of SKA-Mid-AA4 at 1.4 GHz with a 100MHz bandwidth of the left panel. The source is assumed to be at the distance of Fornax A (20 Mpc). Levels increase as $5\sigma\times2^n$, where the reached r.m.s. is $\sigma=2.9\mu$Jy beam$^{-1}$  {\em Right panel}: Same as in the centre panel, but with source located at the distance of \mbox{IC~5063}.}
%     \label{fig:contSim}
% \end{figure}

\section{Jet-ISM interaction: an SKA perspective}
\label{sec:jet_ism_ska}

%\section{Low-redshift studies}
%\subsection{The multiphase gas in nearby AGN} 

%HI + H2(ALMA and JWST) + ionised gas + radio continuum)

\begin{figure}[tbh]
    \centering
	\includegraphics[width=0.9\columnwidth]{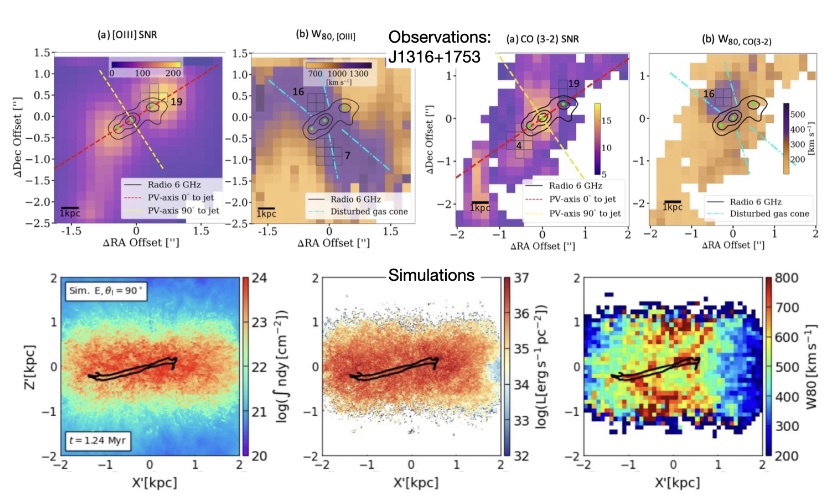}
    \caption{Jet-ISM interaction. The top panels~\citep[from][]{Girdhar2022} show the impact of small-scale jets on the ISM as observed in a $z=0.15$ type-2 quasar. The velocity line width (W80) of the ionized gas (second panel) and cold molecular gas (fourth panel) trace regions of high dispersion (purple spaxels) orthogonally to the jet axis (black contours). The bottom panels~\citep[][]{Meenakshi2022} show the predicted [OIII] emission and line widths (W80) from simulations of jet–ISM interaction, with enhanced widths perpendicular to the jet (black contours). Credit to~\citet{Mukherjee2025}, reproduced with permission.}
    \label{fig:feedbackLowz}
\end{figure}

Relativistic jets from AGN are an important channel for feedback in galaxies. Traditionally, they have been invoked to prevent the cooling of large amount of gas from the circum-galactic scales~\citep{Ciotti2010,OSullivan2011,Fabian2012,McNamara2012}. In the past decades, there has been a rising attention to the impact that these jets may have on galaxy scales (see Fig.~\ref{fig:bwhSmall}), when coupling efficiently with the interstellar medium~\citep[][]{Bicknell1984,Wagner2012,Mukherjee2018,Cielo2018,Young2025}, given the increasing observational evidence of jet-ISM interactions~\citep[see, for example,][]{Combes2013,garcia-burillo2014,Morganti2015,Mahony2016,Santoro2016,Venturi2021,Murthy2022,Audibert2025}. We refer the reader to~\citet{Mukherjee2025} for the most up to date review on the development of numerical simulations to study the impact of AGN driven jets on galaxy scales, which also provides a detailed list of all the confirmed or candidate (sub-)kpc scale jet-ISM interactions that have been currently observed. Two key parameters that seem to regulate the coupling between the jet and the ISM are 1) orientation and 2) jet power. If the jet subtends a small angle relative to the surrounding gaseous disc, then it has a larger cross section with the ISM and it is able to transmit more efficiently its energy to the surrounding gas. Similarly, if the jet has intermediate-power ($P_{\rm jet}<10^{43-44}$ erg s$^{-1}$), it may be trapped for longer time inside the galaxy disc and therefore disrupt the surrounding gas for a longer period, compared to more powerful jets ($P_{\rm jet}<10^{45}$ erg s$^{-1}$) which would more easily and quickly pierce through the ISM and propagate outside the galactic body. A nice example of observational and theoretical evidence of jet-ISM interaction is shown in Fig.~\ref{fig:feedbackLowz}~\citep{Girdhar2022}. The top panels show the kinematics of the ionized gas (T$\sim 10^4$ K) as traced from the [OIII]$\lambda 5007$ line and the cold molecular gas as traced by CO(3-2) in a z$-0.15$ type-2 quasar (J1316+1753). This AGN has a low-power jet detected from Jansky Very Large Array (JVLA) observations (black contours in the figure). The kinematics of the ionized gas clearly shows regions with increased velocity dispersion in a direction perpendicular to the jet axis. This is supported by a dedicated simulation of jet-ISM  interactions (bottom panel) where larger velocity dispersion in [OIII] is predicted perpendicular to the jet. 

This example highlights the importance of combining sensitive radio continuum observations with spatially resolved multi-phase gas components of the ISM. 

%Need to cite something more recent than \mbox{IC~5063} shown in the introduction.

%\textcolor{blue}{ The importance of spatially resolved studies of the multiple-gas components of the ISM combined with high resolution and high sensitivity radio continuum surveys. A recent local examples of these studies is ~\citet[][]{Girdhar2022}. Other works to cite for observations~\citet[][]{Audibert2025,Venturi2021}. Other works to cite for theory~\citet{Mukherjee2018,Meenakshi2022,Cresci2023}}

%\begin{itemize}
% \item Start from Filippo's results on Fornax-A, full resolution
% \item Dive into the simulations a la Dipanjan, starting from the work that we can do now (Girdhar, Da Silva, etc..)
% \item Show what SKA can do, e.g. Filippo will make the predictions from Dipanjan simulations for SKA at different redshfits (where we can also not fully resolving the radio continuum but at least we can get flux and jet axis). 
% \item Isabella should send us prediction of the source populations of jets as a function of redshift with SKA

%\end{itemize}

\subsection{Synthetic observations of AGN feedback with SKA-AA4}

SKAO will make a step forward into the study of jet-ISM interaction enabling the direct comparison of the observed jets in nearby galaxies with simulations at high resolution, thus understanding how the expansion of the relativistic particles through the clumpy ISM of galaxies changes its physical conditions. As illustrated in Figure~\ref{fig:sim_ic5063} (Sect.~\ref{sec:histudies}), so far the comparison between the observed radio jets of \mbox{IC~5063} and the simulated jets of similar power has been limited by the poor resolution of the available 8 GHz observations, which make impossible to measure how differently the jets impact on the gas along the direction of their expansion and perpendicularly to it. In fact, 1.4 GHz observations of \mbox{IC~5063} reveal extended low-power radio lobes~(see Fig.~\ref{fig:ic5063}, left) similar to the ones of \mbox{NGC~1371} (Fig.~\ref{fig:ngc1371}). So far, it has not been possible to understand if these bubbles contributed in unsettling the gas over the galactic scales. 

\begin{figure}[tbh]
    \centering
\begin{subfigure}[b]{0.72\textwidth}
	\includegraphics[width=\columnwidth]{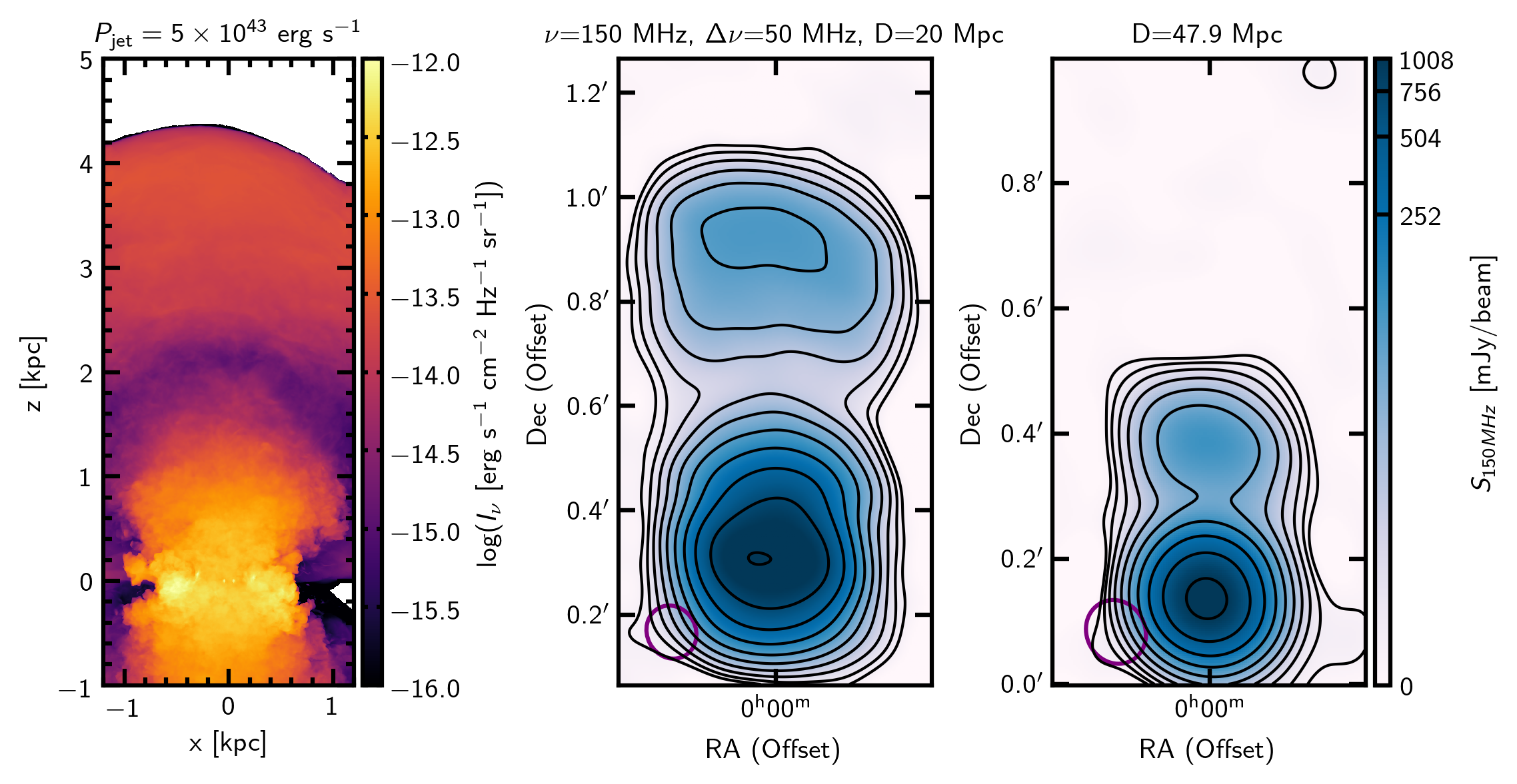}
  % \caption{}
  \label{fig:contSim_low} 
\end{subfigure}
% \medskip % insert a bit of vertical whitespace
\begin{subfigure}[b]{0.72\textwidth}
    \includegraphics[width=\columnwidth]{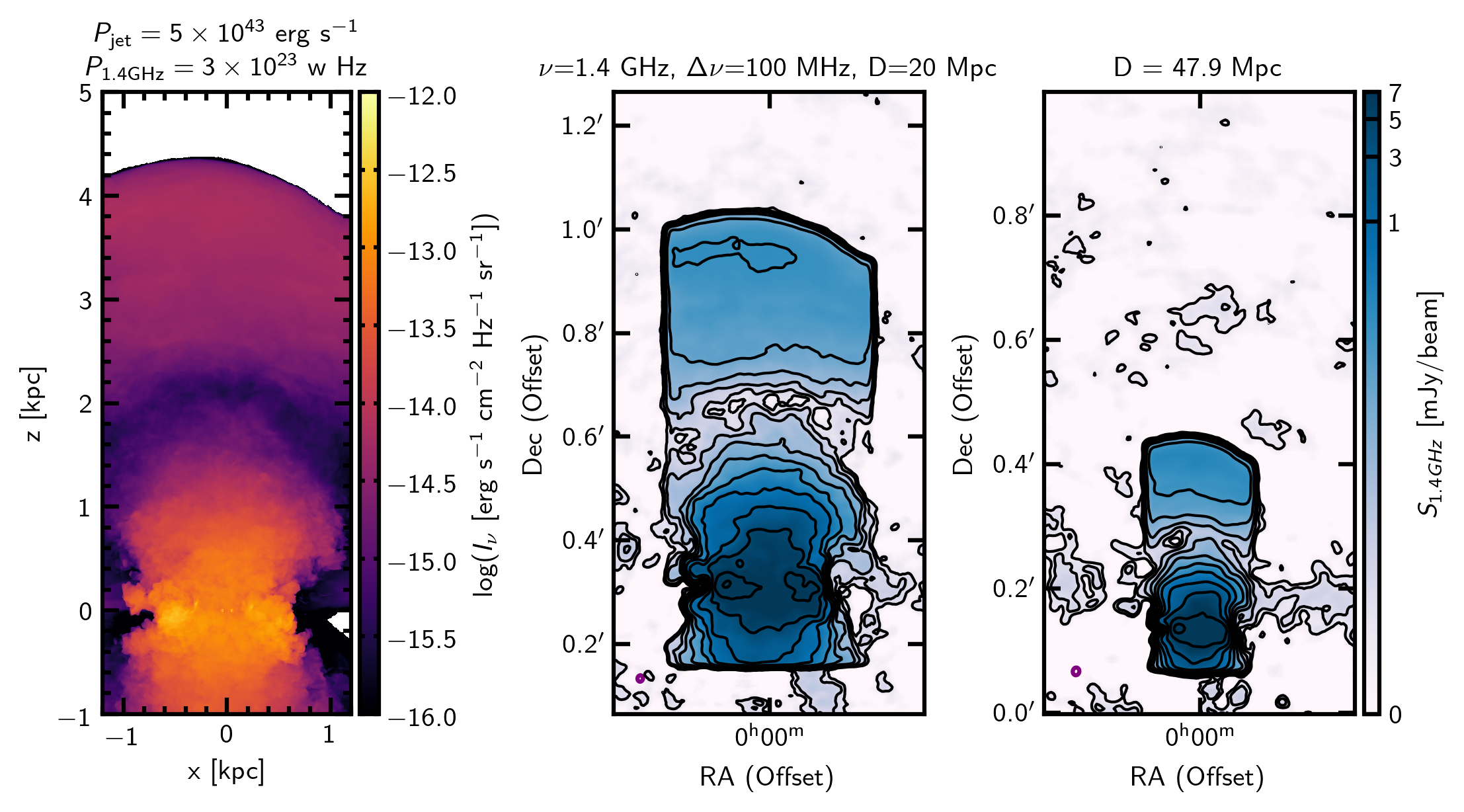}
  % \caption{}
  \label{fig:contSim_high}
\end{subfigure}
\caption[]{{\em Top Left Panel}: Simulation of 300~MHz synchrotron surface brightness produced by the expansion of a low-power jet through a clumpy medium (courtesy of D. Mukherjee). This simulation replicates the radio jets and lobes of \mbox{IC~5063}~\cite[][]{Morganti1998,Oosterloo2000,Mukherjee2018}. {\em Top Centre Panel}: Synthetic observation with 10hrs of SKA-Low-AA4 at 300 MHz with a 85MHz bandwidth of the left panel. The source is assumed to be at the distance of Fornax A (20 Mpc). Levels increase as $5\sigma\times2^n$, where the reached r.m.s. is $\sigma~\sim60\mu$Jy beam$^{-1}$. {\em Top Right Panel}: Same as in the centre panel, but with source located at the distance of \mbox{IC~5063}. {\em Bottom Left Panel}: Simulation of 1.4 GHz synchrotron surface brightness produced by the expansion of a low-power jet through a clumpy medium (Shende et al. in prep, courtesy of D. Mukherjee). This simulation replicates the radio jets and lobes of \mbox{IC~5063}~\cite[][]{Morganti1998,Mukherjee2018} {\em Bottom Centre Panel} Synthetic observation with 10 hours of SKA-Mid-AA4 at 1.4 GHz with a 100MHz bandwidth of the left panel. The source is assumed to be at the distance of Fornax A (20.9 Mpc). Levels increase as $5\sigma\times2^n$, where the reached r.m.s. is $\sigma=0.7\mu$Jy beam$^{-1}$.  {\em Bottom Right Panel}: Same as in the centre panel, but with source located at the distance of \mbox{IC~5063}.}
    \label{fig:contSim}
\end{figure}

Detailed hydrodynamical simulations of low-power jets ($P_{1.4\, \rm GHz}=3\times10^{23}$~\whz) predict that, depending on the jet power and inclination and clumpyness of the medium, synchrotron bubbles may form perpendicularly to the jet and entrain surrounding gas clouds out of the circum-nuclear regions. The left panels of Fig.~\ref{fig:contSim} show a snapshot of the synchrotron emission of these simulations~(Shende et al. in prep) at 300 MHz and 1.4 GHz, respectively. The radio jets expand in the E-W direction for 2 kpc while radio bubbles form in the N-S direction. The central and right panels of Fig.~\ref{fig:contSim} show how SKA-Low and Mid will observe this source. Similarly to what done for the \HI\ in Sect~\ref{sec:cca_synt}, from the simulation we generated the visibilities as the AA4 arrays would observe the synchrotron emission of these jets in 1 hour, at 300 MHz and 1.4 GHz with a bandwidth of 85 and 350 MHz, respectively. Following a data reduction strategy tuned for continuum imaging (robust Briggs $= -1$, tapering $= 0.3''$) we deconvolved the simulated observation and generated the images shown in Fig.~\ref{fig:contSim}. The central panels show the observations of these jets at the distance of Fornax A (20.9 Mpc) while the right panels show the simulated observations at the distance of IC 5063 (47.9 Mpc). Figure~\ref{fig:contSim} demonstrates that SKAO will resolve the full radio structure of these low-power AGN, thus enabling us to directly study their feedback. SKA-Low 300 MHz observations will provide the best trade-off between sensitivity and the resolution required to enable spectral index studies across all SKAO radio bands, $\sim 3''$. Such studies will be crucial to determine the age of the jets enabling us, with the help of the hydrodynamical simulations, to pinpoint how through the different stages of their evolution (active, non-active, restarted) AGN affect physical conditions of the ISM of their hosts. 

\section{SKA Survey Requirements \& Expectations}

In light of the considerations made in the previous Sections, we describe an SKA-AA4 survey plan that sets the minimum requirements to solve the multi-scale problem of AGN feeding and feedback, study these phenomena in detail out to high redshift, and relate the jet-ISM interaction with the SF histories of galaxies.

In detail, SKA-AA4 surveys, with commensal multi-wavelength observations at the same resolution (such as the ones ALMA and IFS), will allow us to understand:

\begin{itemize}
    \item what is the fraction of energy in AGN driven outflows, how and where is it released in the galaxy and how does this change based on jet and environment properties
    \item what is the partition of outflows between cold atomic, molecular and ionised gas. 
    \item how efficiently AGN feedback expels the gas off galaxies and maintains it hot in the halo, preventing its cooling and new star formation episodes.
\end{itemize}

The power of SKA is to perform tiered surveys across all radio bands (from 100 MHz to 15 GHz). This enables resolved spectral index studies of the detected sources, and thus to infer the short timescales of the nuclear activities and relate them to the secular evolution of their host galaxies. Hence, our surveys on AGN feeding and feedback are always tiered across all bands, with different observing times, depending on the bands. 

\subsection{Deep survey of nearby galaxies}

Based on Fig.~\ref{fig:sensPlot_ska}, 10 hours of observations with SKA-Mid AA4 will open a new parameter space in the exploration of low-column density \HI\ in nearby ($\lesssim 100$ Mpc) galaxies. This will enable high resolution studies of the jet-\HI\ interactions across the whole galactic disk of AGN and beyond. Observing $\sim 1000$ galaxies below 100 Mpc for 10 hours per target, AGN feeding and feedback phenomena can be studied in a representative sample of AGN of all radio powers ($10^{21}$-$10^{26}$~\whz) and in star-forming galaxies, allowing a direct comparison of the effects of AGN-ISM interaction with the ones of SF. Such a survey could be carried out by SKA-Mid in $\sim 60$ weeks of observations (assuming \HI\ observations will be carried out also during the day, which is so far not recommended with MeerKAT).

\begin{table}[tbh]
	% \centering
	\caption{Nearby ($\lesssim$ 100 Mpc) galaxy survey requirements}
	\label{tab:surveys}
	\begin{tabularx}{0.99\textwidth}{l c c c c X} % four columns, alignment for each
		\hline\hline
		Band, Freq.  & $\Delta \nu$ & Obs. time  & r.m.s. cont & r.m.s. spec. & Beam \\
		Weighting    & chan. width &   target &  [$\mu$\Jyb]  & [$\mu$\Jyb] & [''] \\
		            &  &[hrs] &   &  [$\times10^{20}$\cmsq, 16~\kms] &  \\

        \hline\hline
		Low, 300 MHz  & 85 MHz  & 1  & $57$ & $9\times10^3$ & $2.3$ \\
		\,rob -2            & 5.43  kHz  &  && &  \\
        \hline
        Band 2, 1.4 GHz  & 350 MHz x2  & 10  & $0.689$ & $107$ & $0.8$ \\
		\,rob 0, tap $0.2''$ & 13.4 kHz  &  & &$35$ &  \\
        Band 2, Zoom HI   & 50 MHz  & 10$^\ddagger$  & $-$ & $186$ & $2.7$ \\
		\,rob 0, tap $1''$ & 3.3 kHz  &  & & $2.2$&   \\		
        Band 2, Zoom OH   & 50 MHz  & 10$^\ddagger$ & $-$ & $176$ & $2.2$ \\
        \, rob -1, tap $0.9''$ & 3.3 kHz  &  && &  \\
        \hline
  %       Band 5, 6.5 MHz   & 3.5 GHz  & 1  & $699$ &  $82$ & $0.16$ \\
		% \, rob 0, tap 0.05'' & 13.44 kHz  &  && &  \\
        % Deep-discovery & Deep targetted follow up of 50 targets identified in the wide-continuum at z$\gtrsim$ 0.05 where to perform spectral index analysis between SKA-Low and Mid at 2'' resolution,\medskip\newline and expand in redshift the detailed analysis of feeding \& feedback phenomena done in the wide-continuum surveys while connecting the timescales of AGN to the SF histories of their hosts 50 targets 20 hrs 0.1  $10^{19}$~\cmsq \\
		\hline\\
	\end{tabularx}
    % \tablefoot{{\bf Total integration time for 100 sources = 1100 hours}. Including $25\%$ overheads, this survey could be completed in less than ten weeks. On top of the targetted sources, this survey will also be a deep survey on random background ($\gtrsim 100$ Mpc) fields, both in continuum and spectral line observation, covering 100 deg$^2$. Spectral line observations should be carried out preferentially at night to avoid solar interference. \\ $\ddagger$: Zoom bands are commensal to Band 2 observations.
\end{table}

The 2'' resolution provided by the AA4 array will enable direct comparison between \HI\ and molecular and ionised gas observations from mm and IFS facilities (such as ALMA and wide-field IFS), across the full stellar body of galaxies. Investigating the low-column density \HI\ regime we will identify the sources of AGN fuelling in different environments (\ie\ diffuse clouds and filaments) as well as resolve in emission the outflowing \HI. The sensitivities and resolutions required to investigate in detail AGN feeding and feedback phenomena in nearby galaxies are summarized in Table~\ref{tab:surveys}.

10 hours observing time at 1.4 GHz will also provide the deepest Band-2 continuum images of nearby AGN. This will enable us to investigate jet-ISM interaction down to very low radio powers ($10^{21}$~\whz), discovering several low-surface brightness radio bubbles and enabling the study of their interaction with the ISM and IGM. These bubbles are predicted by simulations of jet expansion through a clumpy disk (see Fig~\ref{fig:feedbackLowz}) and, besides the ones observed in \mbox{NGC~1371} and \mbox{IC~5063}, have been detected in a handful of objects (see Sect.~\ref{sec:ngc1371}). It is also still a puzzle why extended radio galaxies seem to preferentially be hosted by ellipticals, not spirals~\citep[\eg][]{Kaviraj2015}. This could be a sensitivity issue and the SKAO will discover many more radio jets in spirals~\citep[see some model predictions in Figure 9 of][]{Shabala2017}. If the sources that we think of as compact are actually extended, the SKAO will reveal that AGN feedback over the galactic scales is more efficient than we previously thought.

Deep Band 2 observations of nearby galaxies to study AGN feeding and feedback phenomena are strictly commensal with targetted \HI\ surveys of nearby galaxies which have similar requirements in terms of observing time, spectral and spatial resolution ($\lesssim 1$~\kms, $\sim 2''$), and investigate the connection between \HI\ and star formation, thus revealing in detail the baryon cycle of galaxies~\citep[][]{Rosolowsky01.2026.SKA}, understand how the environment affects the evolution of galaxies and their gas content~\citep[][]{Ramatsoku01.2026.SKA} and search for low-surface brightness dwarf \HI\ galaxies~\citep[][]{Deg01.2026.SKA}. It is worth noting that several nearby SF galaxies show Seyfert-like AGN activity with radio jets~\citep[e.g. the MAGNUM sample][]{Cresci2015,Mingozzi2019,Venturi2021} and are part of the SF nearby galaxy sample of PHANGS~\citep[][]{Schinnerer2019}. Hence, in these galaxies it will be particularly interesting to study the \HI\ with resolution comparable to the ALMA and MUSE observations (currently available for the PHANGS sample) to discriminate the different impacts of AGN and SF activity. In galaxies below 10 Mpc, SKA-Mid AA4 will reach sub-kilo parsec resolution and resolve star-forming regions~\citep[][]{Rosolowsky01.2026.SKA} enabling us to understand how the AGN can change their physical conditions.

Commensal zoom-mode 1.6 GHz observations, will enable sensitive observations of OH-masers with sensitivities sufficient to detect kilo and sub-kilo masers. This will allow us to connect the \HI\ detected in the circum-nuclear regions with the most energetics regions of AGN, in the real proximity ($\lesssim 100$ pc) of the SMBH~\citep[][]{Tarchi01.2026.SKA}.

Tiered SKA-Low 300 MHz observations will be crucial to describe the correct timescales of nuclear activity of nearby AGN. Moreover they will be sensitive to the old relics of past AGN activity, thus revealing how the SMBH may have influenced the galaxy and its IGM also in the past, allowing a direct connection with the SF histories of their hosts. The central frequency of 300 MHz is required to have a PSF comparable to Band 2 observations ($\sim 3''$) which is crucial to produce resolved spectral index maps of the radio jets. As shown in Fig.~\ref{fig:feedbackLowz}, SKA-Low AA4 will allow us to resolve the jets and low-surface brightness lobes of AGN like \mbox{IC~5063} in 1 hour of observation. 

Additional tiered Band 5 observations can provide unique high resolution observations of the jets of nearby galaxies ($0.01 - 0.1''$) and complete the information on the spectral energy distribution of the AGN radio components. For a thorough spectral index analysis, and thus a correct determination of the timescales of nuclear activity it is crucial to correctly identify the frequency breaks of the distribution, which for active or recently switched off nuclei lies at $\gtrsim 2$ GHz. 

The complete characterization of the radio SED of AGN is fundamental and enable the use ofradio jets as probes of the multi-phase ISM, through free-free absorption~\citep{Bicknell2018,Young2025}. The low-frequency turnover of radio SED in compact sources and how this changes with size~\citep[][]{ODea1997} is well explained by a model in which the jet clears out the absorbing material (and hence reduces the free-free absorption optical depth) as it drills through the surrounding multi-phase ISM. Recently,~\citet{Young2025} showed that AGN radio SED is invariant to jet power – it only depends on gas density. Hence, it is possible to study feedback and quantify masses of gas in galaxies using radio data only: combining the full radio SED, the size of the jets (either resolved by SKA-AA4 or by VLBI mode observations) and redshift.

On top of that, observations across all SKA-AA4 bands will exploit commensality with polarization surveys on nearby galaxies~\citep[as discussed in ][]{Mao01.2026.SKA,Tabatabaei01.2026.SKA}. 

A natural follow-up of a deep survey of nearby galaxies will be to probe at high resolution the real proximity of the SMBHs. Resolving against the radio-jets the inflowing and outflowing clouds detected in absorption within the circum-nuclear disk of nearby AGN, it will be possible to directly connect the accretion onto the SMBH and its efficiency, with the properties of the cold outflowing material in the right proximities of the SMBH ($\lesssim 100$ pc), thus completing our knowledge of the feeding and feedback loop. SKA-VLBI observations will be fundamental for these probes~\citep[see][]{Pandey-Pommier03.2026.SKA}.

\subsection{Wide-field continuum survey of galaxies at $z>0.1$}

For AGN feeding and feedback studies it will be crucial to expand the knowledge obtained in the limited sample of nearby galaxies to a larger sample at higher redshifts where AGN feedback is thought to be more efficient in clearing gas out of galaxies. 

\begin{figure}[h]
    \centering
	\includegraphics[width=\columnwidth]{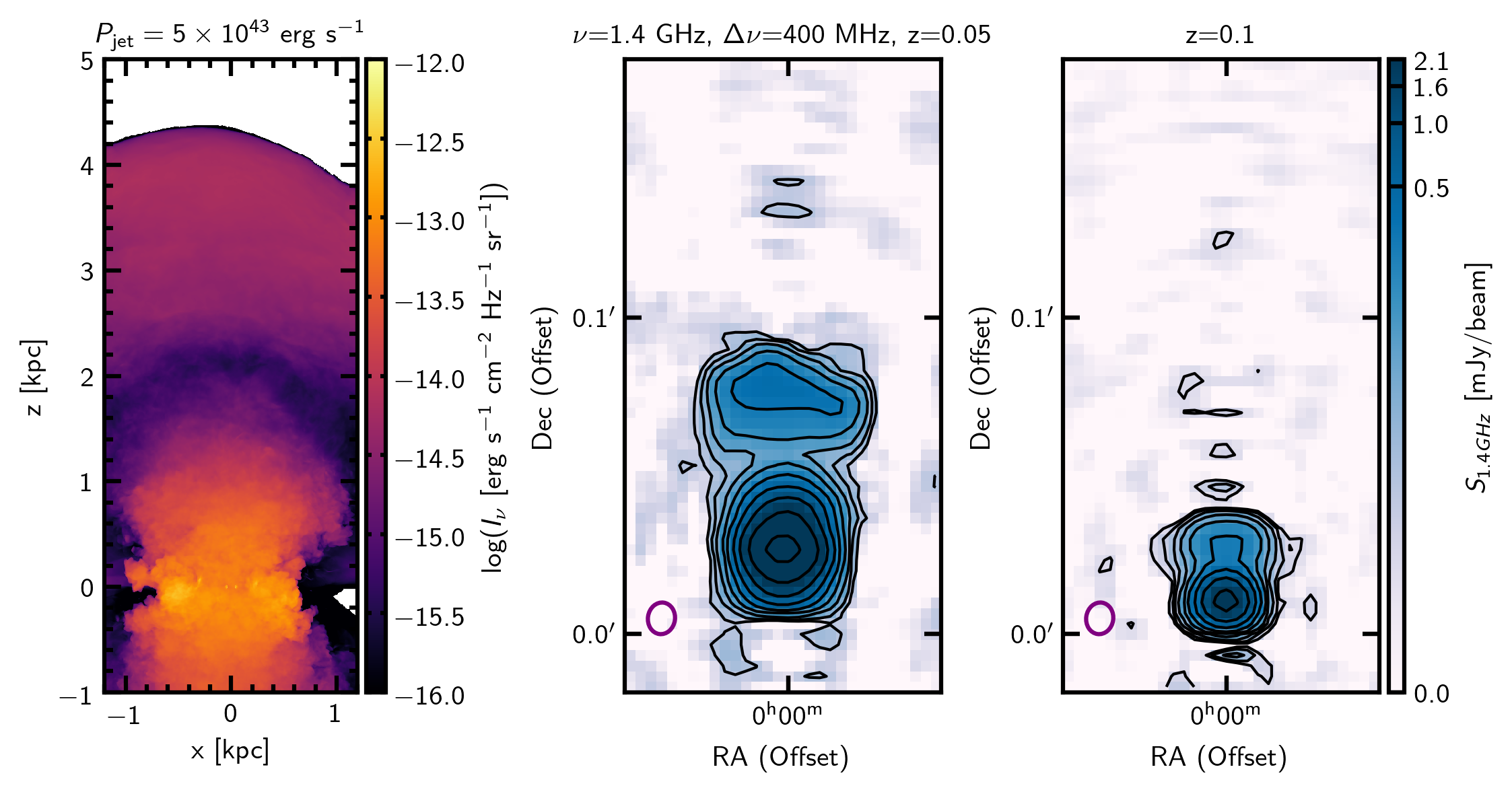}
    \caption{{\em Left Panel}: Simulation of 1.4 GHz synchrotron surface brightness as in Fig~\ref{fig:contSim}. {\em Centre Panel} Synthetic observation of the source at 220 Mpc (z=0.05) with 1.4 hrs of SKA-Mid-AA4 at 1.4 GHz with a 400 MHz. Levels increase as $5\sigma\times2^n$, where the reached r.m.s. is $\sigma=1.2\mu$Jy beam$^{-1}$. {\em Right Panel}: Same as in the centre panel, but with source located at z=0.1, beyond which the 2-kpc jets are not resolved anymore with a psf of 1''.}
    \label{fig:feedback_high}
\end{figure}

The nearby galaxies survey will provide highly detailed diagnostics on the jet-ISM interactions. These will be fundamental to interpret the lower-resolution data of wide continuum surveys which will detect hundreds of radio-AGN out to redshift 5~\citep[see][]{Prandoni01.2026.SKA}. In these surveys, it will be important to resolve the radio jets over the galactic scales to enable jet-ISM studies with the molecular and ionised gas phase inferred with other telescope facilities (see Fig.~\ref{fig:feedbackLowz}). Combining such resolved \HI\ and ionised-gas velocity fields and line-width maps with X-ray spectroscopic constraints on the hot phase will allow us to quantify the turbulent `weather' in group and cluster haloes and test CCA predictions for the link between turbulence, condensation, and jet power (\eg \citealt{Wittor2020,Roncarelli2018}). 

A wide field continuum survey will also allow us to use the statistics of compact AGN jets as a probe of feeding mechanisms. Using sensitive LOFAR samples of active and remnant radio galaxies~\citep[][]{Jurlin2020}, ~\citet{Turner2015} placed strong constraints on the distribution of jet lifetimes and to predict the observable fraction of remnant AGN, and showed that the overall population of radio galaxy jets has a pink noise-like power spectrum in age, consistent with CCA. Because of the high statistics and the constraints set by the number of compact radio sources (where the jets are embedded within the galaxy on kpc-scales), it has been possible to constrain the feeding mechanisms~\citep[][]{Shabala2020}.

Here we provide the requirements that a wide continuum survey should meet to enable AGN feeding and feedback studies over the galactic scales between z=0.05 and $\gtrsim 1$.

\begin{figure}[tbh]
    \centering
	\includegraphics[width=0.45\columnwidth]{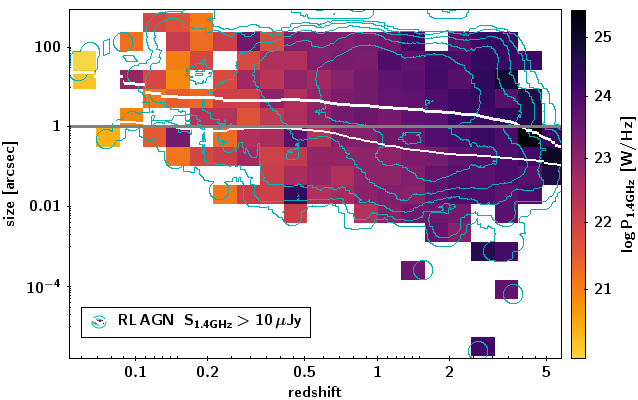}
	\includegraphics[width=0.45\columnwidth]{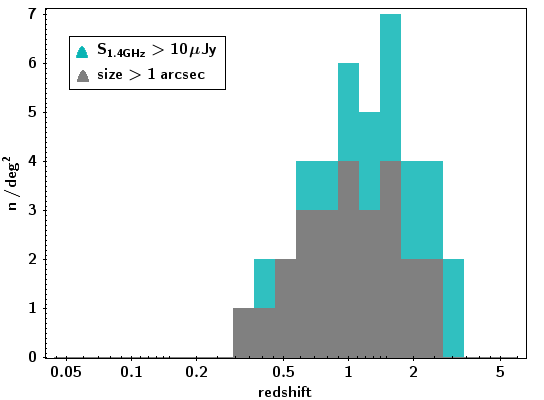}
    \caption{{\em Left Panel}: Distribution of radio AGN on the redshift-size plane (cyan contours), color-coded by radio power at 1.4 GHz for sources detected down to a flux limit of $S_{\rm 1.4 \; GHz}=10\, \mu$Jy ($5\sigma$). The lower white line show the 25\% percentile of the distribution (i.e. 75\% of the sources are above it); the upper white line shows the median of the distribution. The grey line indicate the $1''$ source size. It is clear that $\geq 50\%$ of the sources are resolved at $1"$ resolution, easily achievable with SKA-Mid AA4, at all redshifts, and 75\% are resolved up to $z\sim 0.5$. {\em Right Panel}: Expected redshift distribution of sources with flux density $S_{\rm 1.4 \; GHz}>10\mu$Jy per square degree (cyan histogram). Highlighted in grey is the distribution of the sources larger than 1 arcsec. Distributions taken from the T-RECS simulated catalogues (\citealt{bonaldi23}).}
    \label{fig:contSources}
\end{figure}

In Band 2, a wide-continuum survey with 1 hour of integration time per square degree will reach noise levels of 2 $\mu$\Jyb\ (over 2 sub-bands 350 MHz wide, with robustness 0 and tapering 0.3'' obtaining a PSF of $0.8''$). Figure~\ref{fig:feedback_high} shows that SKA-Mid will resolve both the jets and counter lobes of AGN with moderate radio power, similar to \mbox{IC~5063}, out to $z=0.1$. At higher distances Figure~\ref{fig:contSources} shows that a large fraction ($\geq 50\%$) of the radio AGN that the SKA-Mid will detect with 1 hour observations  at a $5\sigma$ significance ($S_{\rm 1.4 GHz}\gtrsim 10\mu$Jy) are resolved up to redshift $5$ (i.e. their size is larger than 1 arcsec). Between $z=0.1$ and $z=1$ it will be possible to perform resolved studies of the jet-ISM interaction in low and moderate power AGN ($10^{21-23}$~\whz), typically hosted by main sequence galaxies. Figure~\ref{fig:contSources} (b) shows the expected number of sources per square degree, highlighting the need for a wide survey to collect a representative sample of (resolved) radio AGN. 

Observations across all SKA-AA4 bands are crucial for a complete characterisation of the AGN activities. Hence, AGN feeding and feedback studies over the galactic scales will find great benefit from a wide-continuum survey described in detail in~\citet{Prandoni01.2026.SKA}. The main requirement will be to to obtain $\lesssim 3''$ resolution across all bands for resolved spectral index studies. SKA-Low observations centred $300$ MHz over $85$ MHz will reach this requirement in 1 hour of observation. 

% As described in Prandoni et al., tiered Band 1 and Band 2 observations will enable to ....
One hour of observation in Band 2 will enable to detect the \HI\ of SF disks of galaxies ($\sim 10^{21}$~\cmsq), while Band 5 observations will enable observations of $H_2O$ megamasers at high redshifts~\citep[see][]{Tarchi01.2026.SKA}. 

In Band 2, assuming to survey the sky with an hexagonal grid spacing with minimum distance for Nyquist sampling, it will be possible to survey 10000 deg$^2$ with $\sim 3433$ pointings (assuming a primary beam FWHM of 1 degree). By integrating $0.51$ hours per pointing, SKA-Mid will reach the sensitivities of 1 hour per single pointing ($\sim 2.1\mu$\Jyb at $0.85''$ over a bandwidth of $350$ MHz) needed to detect and resolve the jets of sources $\gtrsim 10\mu$~\Jyb (see Fig.~\ref{fig:contSources} (a)), for a total observing time in Band 2 of 1751 hours. Assuming similar observing times for the other SKA-Mid bands and for SKA-Low observations, it will be possible to perform a wide continuum survey over 10000 deg$^2$ in less than a year (41 weeks). Complete knowledge on the stellar properties of the AGN hosts and their ISM will be provided by high resolution multi-plex spectroscopic surveys of facilities like 4MOST. When SKA-AA4 will be start its observations, possibly, also new wide-field ($\sim 1$ deg$^2$), high resolution (R$\sim 40000$) multi-object spectrographs and wide-field ($\sim 10'$-wide) IFS will be available, enabling studies of the multi-phase ISM. 

By combining SKA-AA4 \HI\ and continuum data with these high-resolution spectroscopic constraints, we will be able to test the CCA framework, directly linking cooling and condensation in the halo to SMBH fuelling and the subsequent jet-driven feedback cycle across galactic scales. Moreover, only at these resolutions we will be able to probe the shallow wings of the outflowing ionised gas over multiple emission lines, thus inferring the efficiency of the IGM in shocking the gas, link the jet expansion to the properties of the ISM, such as, for example, temperature, density, metallicity, and enable a direct comparison of the jet-ISM interaction with the diagnostics developed on the nearby galaxy survey.

\paragraph{Acknowledgments} --- FMM carried out part of the research activities described in this paper with contribution of the Next Generation EU funds within the National Recovery and Resilience Plan (PNRR), Mission 4 - Education and Research, Component 2 - From Research to Business (M4C2), Investment Line 3.1 - Strengthening and creation of Research Infrastructures, Project IR0000034 – “STILES - Strengthening the Italian Leadership in ELT and SKA”.
M.G. acknowledges support from the ERC Consolidator Grant \textit{BlackHoleWeather} (101086804).
WJGdB: This work has received funding from the European Research Council
(ERC) under the European Union’s Horizon 2020 research and innovation
programme (grant agreement No. 882793 ``MeerGas'').
SSS acknowledges funding from the Australian Research Council via grant DP240102970.

\bibliographystyle{abbrvnat-maxbibnames4}
\bibliography{AGN_FF} % if your bibtex file is called example.bib

\end{document}

%% file: journal-names.tex
\newcommand{\actaa}{Acta Astron.} % Acta Astronomica
\newcommand{\araa}{ARA\&A} % Annual Review of Astron and Astrophys
\newcommand{\aar}{A\&ARv} % Astrononmy \& Astrophysics Review
\newcommand{\aapr}{A\&ARv} % Astronomy\&Astrophysics Reviews
\newcommand{\ab}{Astrobiol.} % Astrobiology
\newcommand{\aj}{AJ} % Astronomical Journal
\newcommand{\apj}{ApJ} % Astrophysical Journal
\newcommand{\apjl}{ApJL} % Astrophysical Journal, Letters
\newcommand{\apjs}{ApJSS} % Astrophysical Journal, Supplement
\newcommand{\ao}{Appl. Opt.} % Applied Optics
\newcommand{\apss}{Astro. \& Space Sci.} % Astrophysics and Space Science
\newcommand{\aap}{A\&A} % Astronomy and Astrophysics
\newcommand{\aaps}{A\&AS.} % Astronomy and Astrophysics, Supplement
\newcommand{\baas}{Bull. Am. Astron. Soc.} % Bulletin of the AAS
\newcommand{\caa}{Chinese A\&A} % Chinese Astronomy and Astrophysics
\newcommand{\cjaa}{Chinese J. A\&A} % Chinese Journal of Astronomy and Astrophysics
\newcommand{\cqg}{Class. Quantum Gravity} % Classical and Quantum Gravity
\newcommand{\gal}{Galaxies} % Galaxies
\newcommand{\gca}{Geo. Cosmo. Acta} % Geochimica Cosmochimica Acta
\newcommand{\icarus}{Icarus} % Icarus
\newcommand{\jcap}{JCAP} % Journal of Cosmology and Astroparticle Physics
\newcommand{\jgr}{J. Geophys. Res.} % Journal of Geophysics Research
\newcommand{\jgrp}{J. Geophys. Res. Planets} % Journal of Geophysics Research: Planets
\newcommand{\jqsrt}{J. Quant. Spectrosc. Radiat. Transf.} % Journal of Quantitiative Spectroscopy and Radiative Transfer
\newcommand{\memsai}{Mem. SAIt} % Mem. Societa Astronomica Italiana
\newcommand{\mnras}{MNRAS} % Monthly Notices of the RAS
\newcommand{\nat}{Nature} % Nature
\newcommand{\nastro}{Nat. Astron.} % Nature Astronomy
\newcommand{\ncomms}{Nat. Commun.} % Nature Communications
\newcommand{\nphys}{Nat. Phys.} % Nature Physics
\newcommand{\na}{New Astron.} % New Astronomy
\newcommand{\nar}{New Astron. Rev.} % New Astronomy Review
\newcommand{\physrep}{Phys. Rep.} % Physics Reports
\newcommand{\pra}{Phys. Rev. A} % Physical Review A: General Physics
\newcommand{\prb}{Phys. Rev. B} % Physical Review B: Solid State
\newcommand{\prc}{Phys. Rev. C} % Physical Review C
\newcommand{\prd}{Phys. Rev. D} % Physical Review D
\newcommand{\pre}{Phys. Rev. E} % Physical Review E
\newcommand{\prx}{Phys. Rev. X} % Physical Review X
\newcommand{\prl}{Phys. Rev. Let.} % Physical Review Letters
\newcommand{\psj}{Planet. Sci. J.} % Planetary Science Journal
\newcommand{\planss}{Planet. Space Sci.} % Planetary Space Science
\newcommand{\pnas}{Proc. Natl Acad. Sci. USA} % Proceedings of the US National Academy of Sciences
\newcommand{\procspie}{Proc. SPIE} % Proceedings of the SPIE
\newcommand{\pasa}{PASA} % Publications of the Astron.  Soc. of Australia
\newcommand{\pasj}{PASJ} % Publications of the Astron.  Soc. of Japan 
\newcommand{\pasp}{PASP} % Publications of the Astron.  Soc. of the Pacific
\newcommand{\rmxaa}{RMXAA} % Revista Mexicana de Astronomia y Astrofisica
\newcommand{\sci}{Science} % Science
\newcommand{\sciadv}{Sci. Adv.} % Science Advances
\newcommand{\solphys}{Sol. Phys.} % Solar Physics
\newcommand{\sovast}{Soviet Ast.} % Soviet Astronomy
\newcommand{\ssr}{Space Sci. Rev.} % Space Science Reviews
\newcommand{\uni}{Universe} % Universe